\documentclass[prl,superscriptaddress,showpacs,letterpaper,10pt,twocolumn]{revtex4}
\usepackage{color,graphicx}
\usepackage{bm}
\usepackage{amsmath}
\usepackage{amssymb}
\usepackage{epstopdf}
\usepackage{ulem}
\usepackage{url}

\def\be{\begin{equation}}
\def\ee{\end{equation}}
\def\bea{\begin{eqnarray}}
\def\eea{\end{eqnarray}}

\begin{document}
\title{Magnetic and Ising quantum phase transitions in a model for isoelectronically tuned iron pnictides}
\author{Jianda Wu}
\affiliation{Department of Physics \& Astronomy, Rice University, Houston, Texas 77005, USA}
\affiliation{Department of Physics, University of California, San Diego, California 92093, USA}
\author{Qimiao Si}
\affiliation{Department of Physics \& Astronomy, Rice University, Houston, Texas 77005, USA}
\author{Elihu Abrahams}
\affiliation{Department of Physics and Astronomy, University of California Los Angeles, Los Angeles,
CA 90095, USA}
\begin{abstract}
Considerations of the observed bad-metal behavior in Fe-based superconductors led to an early proposal
for quantum criticality induced by isoelectronic P for As doping in iron arsenides, which has
since been experimentally confirmed. We study here an effective model for the isoelectronically tuned
pnictides using a large-$N$ approach. The model contains antiferromagnetic and Ising-nematic order parameters
appropriate for $J_1$-$J_2$ exchange-coupled local moments on an Fe square lattice,
and a damping caused by  coupling to itinerant electrons.
The zero-temperature magnetic and Ising transitions
are concurrent and essentially continuous. The order-parameter jumps are very small,
and are further reduced by the inter-plane coupling;
consequently,
quantum criticality
occurs over a wide dynamical range.
Our results
reconcile recent seemingly contradictory experimental observations concerning the
quantum phase transition in the P-doped iron arsenides.
\end{abstract}
\maketitle
\paragraph*{Introduction.}Iron pnictide and chalcogenide materials not only show high-temperature
superconductivity \cite{kamihara},
but also
feature rich phase diagrams. For the  undoped parent iron arsenides,
the ground state has collinear $(\pi,0)$ magnetic order \cite{delaCruz}.
Because superconductivity occurs at the border of this antiferromagnetic (AF) order,
a natural question is whether quantum
criticality plays a role in the phase diagram. Early on, it was  proposed theoretically that tuning the
parent iron arsenide by isoelectronic P-for-As doping induces
quantum criticality associated with the suppression of
both the $(\pi,0)$ AF order
and an Ising-nematic spin order \cite{dai}.
This proposal was made
within a strong-coupling approach, which attributes the bad-metal behavior
of iron arsenides \cite{qazilbash,hu,degiorgi,nakajima} to correlation effects that are
on the verge of localizing electrons \cite{si,si2,kotliar2} along with their associated magnetic moments.
The P doping increases the in-plane electronic kinetic energy (as P is smaller than As),
and thus the coherent electronic spectral
weight  while leaving other model parameters little changed \cite{quebe,zimmer}.
This weakens both the magnetic order and the associated Ising-nematic spin order
\cite{dai,abrahams}.

Experimental evidence for a quantum critical point (QCP) has since emerged
in the P-doped CeFeAsO
\cite{cruz,luo} and P-doped BaFe$_2$As$_2$ \cite{kasahara,nakai,hashimoto,analytis,analytis2}.
In the phase diagram of the P-doped BaFe$_2$As$_2$,
an extended temperature and doping regime has been identified
for non-Fermi liquid
behavior
\cite{kasahara,nakai,hashimoto,analytis}.
An Ising-nematic order, inferred from the tetragonal-to-orthorhombic structural distortion,
is suppressed around the same P-doping concentration ($x_c \approx 0.33$) at which
the AF order disappears.
While there is evidence for a QCP ``hidden'' inside the superconducting dome \cite{hashimoto},
quantum criticality has now been observed and
studied in the normal state when superconductivity
is suppressed by a high field \cite{analytis,analytis2}.
We note that the bad-metal behavior persists through $x_c$ \cite{kasahara}.

Recently, evidence for a weakly first-order nature of the transition has come
from the neutron-scattering experiments in the P-doped BaFe$_2$As$_2$ \cite{DHu}.
It is in seeming contradiction with the accumulated experimental evidence for quantum criticality.
This puzzle calls for further theoretical analyses on the underlying quantum phase transitions.
More generally, the interplay between the magnetic and nematic orders exemplifies the kind of
competing or coexisting orders that is of general interest to
a variety of strongly correlated electron systems.

In this letter, we study the zero-temperature phase transitions in the  appropriate effective Ginzburg-Landau
field theory that was
introduced earlier \cite{dai,abrahams} to describe the low-energy properties of a $J_1$-$J_2$ model of local moments
on a square lattice coupled to coherent itinerant electrons \cite{dai,chandra,si,fang,xu}.
The theory
contains antiferromagnetic (vector) and Ising-nematic (scalar) order parameters
as well as a damping term.
Since it is important to establish the nature of quantum criticality in the absence of superconductivity
\cite{cruz,analytis,analytis2},
we will focus on the transitions in the normal state and will
not consider the effect of superconductivity \cite{fernandes}.
Using a large-$N$
approach \cite{rancon,sidney-coleman}, we demonstrate that the
AF and Ising-nematic transitions
are concurrent at zero temperature
both for the case of a square lattice
and in the presence of interlayer coupling.
Moreover, both transitions are
only weakly first order in accordance with the marginal nature of the relevant coupling,
with jumps in both order parameters that are very small,
which implies a large dynamical range for quantum criticality.
Our results provide a natural resolution to the aforementioned puzzle.

\begin{figure*}[t]
\begin{center}
\includegraphics[width=13cm]{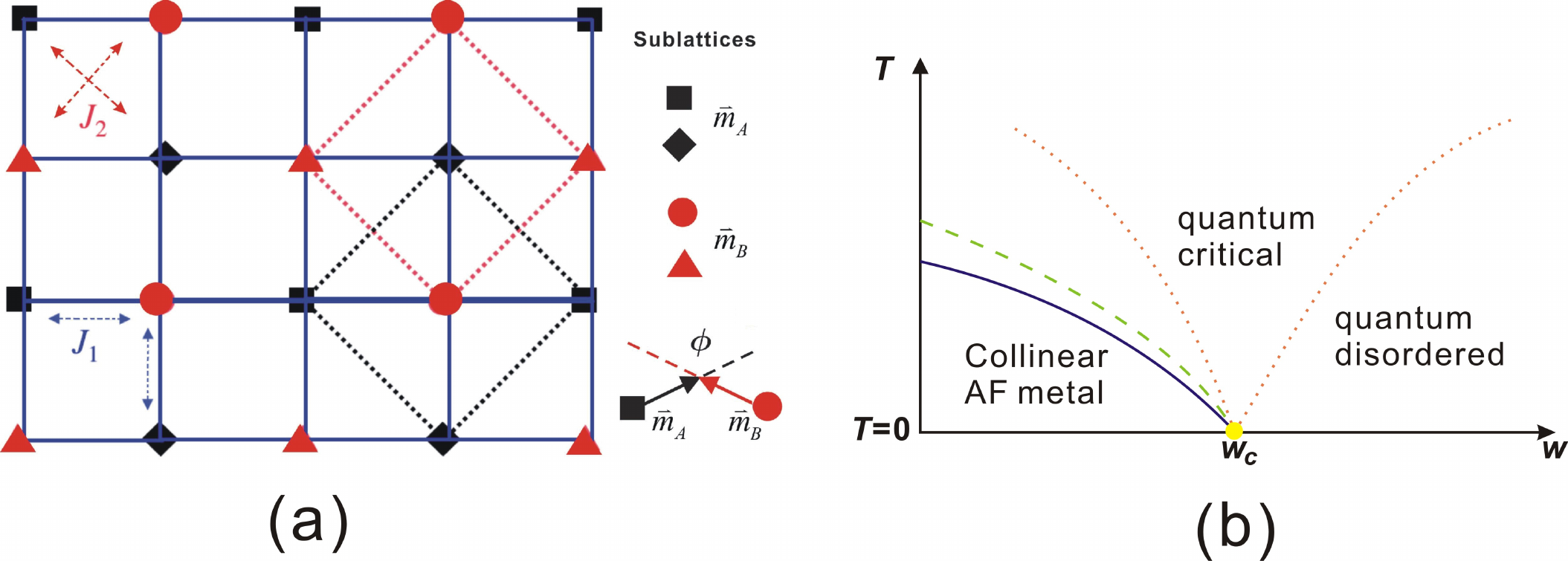}
\end{center}
\caption{(a) Illustration of the $J_1-J_2$ model on a square lattice. The staggered magnetizations
$\overset{\lower0.5em\hbox{$\smash{\scriptscriptstyle\rightharpoonup}$}} {m} _A$ and
 $\overset{\lower0.5em\hbox{$\smash{\scriptscriptstyle\rightharpoonup}$}} {m} _B$ are defined on two interpenetrating
N\'eel square lattices; (b) Schematic phase diagram proposed for the P-doped iron arsenides \cite{dai}.
P-doping increases $w$, the  spectral weight of the coherent itinerant electrons.
The yellow dot denotes the tuning parameter $w_c$ for the QCP. The purple solid line and the green dashed
one respectively mark the AF and structural transitions.}
\label{fig:combine0}
\end{figure*}
\paragraph*{The model.}The proximity of a bad metal to a Mott transition can be measured by a parameter $w$,
the percentage of the single-electron spectral weight in the coherent itinerant part \cite{dai,si,si2,moeller}.
This approach has been successful in describing the
spin excitation spectrum of the iron pnictides \cite{yu,pallab,Diallo2010,Harriger2011},
and in understanding the fact that $T_c$ in the iron-based superconductors with purely electron
Fermi pockets is at least comparably high compared
with those with nested Fermi surfaces of co-existing hole and electron pockets
\cite{RYu2013a,JGuo2010,Fang2011,Xue.2012,SLHe2013,Shen.2014,YWang2015}.
At the zeroth order in $w$, all the single-electron excitations are incoherent; integrating
 out
 the corresponding charge excitations
 leads to couplings $J_1$ and $J_2$
among the residual local moments:
\be
H = \sum\limits_{\left\langle {i,j} \right\rangle } {J_1 \overset{\lower0.5em\hbox{$\smash{\scriptscriptstyle\rightharpoonup}$}}
{S} _i  \cdot \overset{\lower0.5em\hbox{$\smash{\scriptscriptstyle\rightharpoonup}$}} {S} _j }
+ \sum\limits_{\left\langle {\left\langle {i,j} \right\rangle } \right\rangle }
{J_2 \overset{\lower0.5em\hbox{$\smash{\scriptscriptstyle\rightharpoonup}$}} {S} _i
\cdot \overset{\lower0.5em\hbox{$\smash{\scriptscriptstyle\rightharpoonup}$}} {S} _j } \label{model0}
\ee
where $\left\langle {\cdots} \right\rangle$ and
$\left\langle {\left\langle  \cdots  \right\rangle } \right\rangle
$ respectively denote the nearest neighbor and next nearest neighbor sites;
see Fig.~\ref{fig:combine0}(a).
Both general considerations \cite{si} and first-principal calculations \cite{yildirim,ma} suggest
 that $J_2 > J_1/2$. In this regime, we consider
two interpenetrating
sublattices [the dotted squares in Fig.~\ref{fig:combine0}(a)],
having independent staggered magnetizations with N\'eel vectors
$\overset{\lower0.5em\hbox{$\smash{\scriptscriptstyle\rightharpoonup}$}} {m} _A$
and $\overset{\lower0.5em\hbox{$\smash{\scriptscriptstyle\rightharpoonup}$}} {m} _B$.
While the mean-field energy is independent of the angle $\phi$
between $\overset{\lower0.5em\hbox{$\smash{\scriptscriptstyle\rightharpoonup}$}} {m} _A$
and $\overset{\lower0.5em\hbox{$\smash{\scriptscriptstyle\rightharpoonup}$}} {m} _B$,
this degeneracy is
broken by
quantum or thermal fluctuations.
It leads to the collinear
order with $\phi = 0$ or $\pi$ \cite{henley,chandra}.
Thus $\overset{\lower0.5em\hbox{$\smash{\scriptscriptstyle\rightharpoonup}$}}
{m} _A\cdot \overset{\lower0.5em\hbox{$\smash{\scriptscriptstyle\rightharpoonup}$}}
{m} _B = \pm 1$ becomes an Ising variable.

At non-vanishing orders in $w$, the coherent itinerant electrons provide Landau damping.
This leads to the following Ginzburg-Landau action \cite{dai,abrahams}:
\be
S = S_2  + S_4
\ee
with
\begin{widetext}
\be
\begin{gathered}
S_2  = \sum\limits_{\overset{\lower0.5em\hbox{$\smash{\scriptscriptstyle\rightharpoonup}$}} {q} ,i\omega _l }
{\left\{ {\chi _0^{ - 1} (\overset{\lower0.5em\hbox{$\smash{\scriptscriptstyle\rightharpoonup}$}} {q} ,i\omega _l )
\left[ {\left| {\overset{\lower0.5em\hbox{$\smash{\scriptscriptstyle\rightharpoonup}$}} {m} _A
(\overset{\lower0.5em\hbox{$\smash{\scriptscriptstyle\rightharpoonup}$}} {q} ,i\omega _l )} \right|^2
+ \left| {\overset{\lower0.5em\hbox{$\smash{\scriptscriptstyle\rightharpoonup}$}} {m} _B
(\overset{\lower0.5em\hbox{$\smash{\scriptscriptstyle\rightharpoonup}$}} {q} ,i\omega _l )} \right|^2 } \right]
+ 2v\left( {q_x^2  - q_y^2 } \right)\overset{\lower0.5em\hbox{$\smash{\scriptscriptstyle\rightharpoonup}$}}
 {m} _A (\overset{\lower0.5em\hbox{$\smash{\scriptscriptstyle\rightharpoonup}$}} {q} ,i\omega _l )
  \cdot \overset{\lower0.5em\hbox{$\smash{\scriptscriptstyle\rightharpoonup}$}} {m} _B
   ( - \overset{\lower0.5em\hbox{$\smash{\scriptscriptstyle\rightharpoonup}$}} {q} , - i\omega _l )} \right\}}, \hfill \\
  S_4  = \int_0^\beta  {d\tau \int {d\overset{\lower0.5em\hbox{$\smash{\scriptscriptstyle\rightharpoonup}$}} {r}
   \left\{ {u_1 \left( {\left| {\overset{\lower0.5em\hbox{$\smash{\scriptscriptstyle\rightharpoonup}$}} {m} _A } \right|^4
    + \left| {\overset{\lower0.5em\hbox{$\smash{\scriptscriptstyle\rightharpoonup}$}} {m} _B } \right|^4 } \right)
    + u_2 \left| {\overset{\lower0.5em\hbox{$\smash{\scriptscriptstyle\rightharpoonup}$}} {m} _A } \right|^2
    \left| {\overset{\lower0.5em\hbox{$\smash{\scriptscriptstyle\rightharpoonup}$}} {m} _B } \right|^2  - u_I
    \left( {\overset{\lower0.5em\hbox{$\smash{\scriptscriptstyle\rightharpoonup}$}} {m} _A
    \cdot \overset{\lower0.5em\hbox{$\smash{\scriptscriptstyle\rightharpoonup}$}} {m} _B } \right)^2 } \right\}} }.  \hfill \\
\end{gathered}
\label{action}
\ee
\end{widetext}
The ${\vec m}_{A/B}$ are in either momentum and Matsubara frequency space ($S_2$) or real space and imaginary time ($S_4$).
In $S_2$, the inverse susceptibility is
\be
\chi _0^{ - 1} (\mathord{\buildrel{\lower3pt\hbox{$\scriptscriptstyle\rightharpoonup$}}
\over q} ,i\omega _l ) = r + \omega _l^2  + c~q^2  + \gamma \left| {\omega _l } \right|,
\label{chi}
\ee
where $c$ is the square of the spin-wave velocity and in $S_4$,
the coupling $u_I>0$
\cite{dai,chandra}. The parameter $v$ leads to
the anisotropic distribution of the spin spectral weight in momentum space, which is observed
in neutron scattering \cite{Diallo2010,Harriger2011}.
It is described by the ellipticity
\be
\epsilon \equiv \sqrt{(c-v)/(c+v)} ,
\ee
which goes from full isotropy $\epsilon =1$ ($v = 0$) to
extreme anisotropy $\epsilon = 0$ ($v = c$).
In addition, $\gamma$ is the (Landau) damping rate
and $r = r_0 +wA_{\bf Q}$, where $r_0$ is negative, reflecting ground-state order in the absence of damping,
and $A_{\bf Q}>0$ is
related to
a quasiparticle susceptibility at ${\bf Q}=(\pi,0)$ or $(0,\pi)$ \cite{dai}.
The mass $r$ vanishes
at $w=w_c$, the point of quantum phase transition.
When the damping is present, the effective dimensionality of the fluctuations
is $d+z =4$.
From a renormalization-group (RG) perspective, because ``$- u_I$'' is negative, it is marginally relevant
w.r.t the underlying QCP at $d+z=4$ \cite{dai,qi}.
So unlike thermally-driven transitions or the case of a zero-temperature transition
in the absence of damping (where $u_I$ is relevant),
the marginal nature of the coupling is expected to
yield only a small change to the underlying QCP; this leads
to a qualitative phase diagram shown in Fig.~\ref{fig:combine0}(b) \cite{dai,abrahams}.

Given the aforementioned experimental observations,
we shall study the phase transitions
beyond qualitative RG-based considerations.
Our focus is on the zero-temperature limit, and we place
particular emphasis on the effect of damping. We note that the effect of damping on the transitions and dynamics
at non-zero temperatures
has been studied before \cite{pallab}.
The action $S$ is a functional of the (vector) magnetization fields ${\vec m}_{A/B}$
and we may derive the free-energy density from ${\cal F} = -\ln \int{\cal D}\{m\}\exp(-S(\{m\})$.

\paragraph*{Large-$N$ approach.---}
To study  the phase transitions for  the two-sublattice  action of Eq.~(\ref{action})
beyond mean-field theory, we generalize the spin symmetry of the model  to $O(N)$ (${\vec m}_{A/B}$
will have $N$ components) and study it
through a $1/N$ expansion.
Our goal is to investigate general properties, including issues of universality
and the order  of the phase transitions of the present setting, which  contains {\it two} order parameters
possibly competing or coexisting.
We note that
the well-known large-$N$ approach has proved fruitful for many problems in statistical physics
\cite{rancon,sidney-coleman}.

To proceed, we rescale the quartic couplings in $S(\{m\})$ by a factor $1/N$
and  in the functional integral over $e^{-S}$ for ${\cal F}$, we decompose them in terms of Hubbard-Stratonovich fields
$\lambda _{A/B}$ and $\Delta _I.$ For details, refer to the Supplementary Material (SM) \cite{supplemental}.
To leading order in $1/N$,
$i\lambda _{A/B}  =\langle m_{A/B}^2\rangle \equiv m^2$
contribute to the renormalization of the mass
(coefficient of the quadratic term in the action $S_2$)
and
$\; \Delta _I  = \left\langle {\mathord{\buildrel{\lower3pt\hbox{$\scriptscriptstyle\rightharpoonup$}}
\over m} _A  \cdot \mathord{\buildrel{\lower3pt\hbox{$\scriptscriptstyle\rightharpoonup$}}
\over m} _B } \right\rangle $
is the Ising order parameter. We carry out our analysis from
the ordered side, and set $\mathord{\buildrel{\lower3pt\hbox{$\scriptscriptstyle\rightharpoonup$}}
\over m} _{A/B}  = \left( {\sqrt N \sigma _{A/B} ,\mathord{\buildrel{\lower3pt\hbox{$\scriptscriptstyle\rightharpoonup$}}
\over \pi } _{A/B} } \right)$ with ${\sigma _{A/B} }$ and ${\mathord{\buildrel{\lower3pt\hbox{$\scriptscriptstyle\rightharpoonup$}}
\over \pi } _{A/B} }$ as the static order and fluctuation fields of sublattices $A$ and $B$ respectively.
To order  $O(1/N)$ we can integrate out ${\mathord{\buildrel{\lower3pt\hbox{$\scriptscriptstyle\rightharpoonup$}}\over \pi }_{A/B}}$,
which yields an effective free energy density ${\cal F}(\sigma,m^2,\Delta_I)$
that depends parametrically on  the damping  strength $\gamma$, the square of the spin-wave velocity $c$,
the anisotropy parameter $v$, and the  quartic coupling constants $u_I,  2u_1+u_2$.
From SM, Eqs.~(S6,S7), we have the free energy density
\be
{\cal F} = \frac{\Delta _I^2}
{u_I} - \frac{(m^2  - r)^2}
{2u_1  + u_2} + (m^2  \pm \Delta _I )\sigma ^2  + g(m^2,\Delta_I)  \label{free}
\ee
with
\begin {align}
  g(m^2,\Delta_I) &= \frac{1}{2\beta V}\sum_{{\vec q},l} \ln\big\{(D_{0,{\vec q},l}^{-1}  + m^2)^2 \notag\\
   &-[v(q_x^2  - q_y^2) + \Delta _I]^2\big\},
\label{ironpnictides:partoffree}
\end{align}
where $D_{0,{\vec q},l} ^{-1}= \chi^{-1}_{0,{\vec q},l} -r$, containing $\gamma$, see Eq.~(\ref{chi}).
The two cases $\sigma_A = \sigma_B  =\sigma$ ($+$sign in Eq.~(\ref{free})) and $\sigma_A = -\sigma_B = \sigma$ ($-$sign)
correspond to
${\bf{Q}} = (0,\pi)$
and $(\pi,0)$)
AF orders,
respectively.

Then we have variational equations w.r.t $\sigma$, $m^2$ and $\Delta_I$,
\be
\frac{{\partial {\cal F}}}{{\partial \sigma  }} = \frac{{\partial {\cal F}}}{{\partial \Delta _I }} = \frac{{\partial {\cal F}}}{{\partial m^2 }} =0
\label{v1}
\ee
which in turn correspond to [see  SM, Eqs.~(S9-S11)] \cite{supplemental}
\bea
&\left( { m^2 - |\Delta_I| } \right)\sigma  = 0, \label{v2} \\
\frac{{\Delta _I }}{{u_I }} &= \frac{{m^2  - r}}{{2u_1  + u_2 }} - 2 \sigma ^2  - G_+ ,  \label{v3}  \\
 \frac{{\Delta _I }}{{u_I }} &=  - \frac{{m^2  - r}}{{2u_1  + u_2 }} + G_- .  \label{v4}
\eea
Here $G_{\pm}$ are given by
\be
G_{\pm} = \frac {1}{2\beta V}\sum _{{\vec q},l}\frac{1}{ D_{0,{\vec q},l}^{-1}\pm v(q_x^2-q_y^2) +m^2 \pm \Delta_I}.
\ee

Several limits provide a check on our approach.
From Eqs.~(\ref{v3},\ref{v4}), setting $u_I = 0$ will lead to $\Delta_I = 0$; this is consistent with
the Ising order being driven by the interaction $u_I$.  In the absence of coupling to coherent itinerant fermions
i.e., setting  $\gamma^2/|\Delta_I| = 0$ and $w=0$,
we have a nonzero Ising order at zero temperature, which is
what happens for the pure $J_1-J_2$ model
\cite{henley,chandra}.
The detailed analysis of these saddle-point equations is in SM, Eqs.~ (S12-S15).
It follows that the vanishing of the Ising order implies
a vanishing magnetic order. The converse can also be shown explicitly by
analyzing Eq.~(S15) of the SM,
and is numerically confirmed (see below).

\paragraph*{Nature of the magnetic and Ising transitions at zero temperature.---}
We are now in position to address the concurrent magnetic and Ising transition at $T=0$.
The RG argument we described earlier suggests that there will be a jump of the order parameters
across the transition, but the jump will be smaller as the damping parameter $\gamma$ increases.
To see how the damping affects the transition, we first consider the parameter regime where analytical
insights can be gained in our large-$N$ approach. When $\gamma$ is sufficiently large so that
$x, y \ll 1$, Eq.~(\ref{v3}) simplifies to be \cite{supplemental}
\be
A(\eta)= a \eta - \eta\ln \eta = \mu(w) \label{leadingising}
\ee
with $\eta = | \Delta_I | /\gamma ^2 $, and
\bea
\begin{gathered}
  a =  - \frac{{8\pi ^2 \Gamma\left( {a_I  - a_0 } \right)}}
{{\epsilon + 1/\epsilon }} - \ln 2 - 1/2 , \hfill \\
\mu (w) = \frac{{8\pi ^2 a_0 }}
{{(\epsilon+1/\epsilon) \Gamma }} \frac{{r(w)}}
{{c\Lambda _c^2 }} + \frac{\tan ^{ - 1} (2/\Gamma) }
{{\Gamma }} - \frac{1}
{4}\ln (1 + \frac{4}{\Gamma^2}) , \hfill \\
\end{gathered} \label{ironpnictides:a}
 \eea
 where
$ \Gamma  = \frac{\gamma }
{{c^{1/2} \Lambda _c }}$ is the normalized damping rate,
while $a_0 = \frac{{\Lambda _c c^{3/2} }}{{2u_1  + u_2 }}$ and
$a_I = \frac{{\Lambda _c c^{3/2} }}{u_I}$ relate to the normalized interactions.
As described in detail in the Supplementary Material \cite{supplemental},
it follows from this equation that the transition is first order,
with the jump of the order parameter decreasing as the damping rate $\Gamma$ is increased.
The jump is exponentially suppressed when $\Gamma$ becomes large.

\begin{figure}[t!]
\begin{center}
\includegraphics[width=7.5cm]{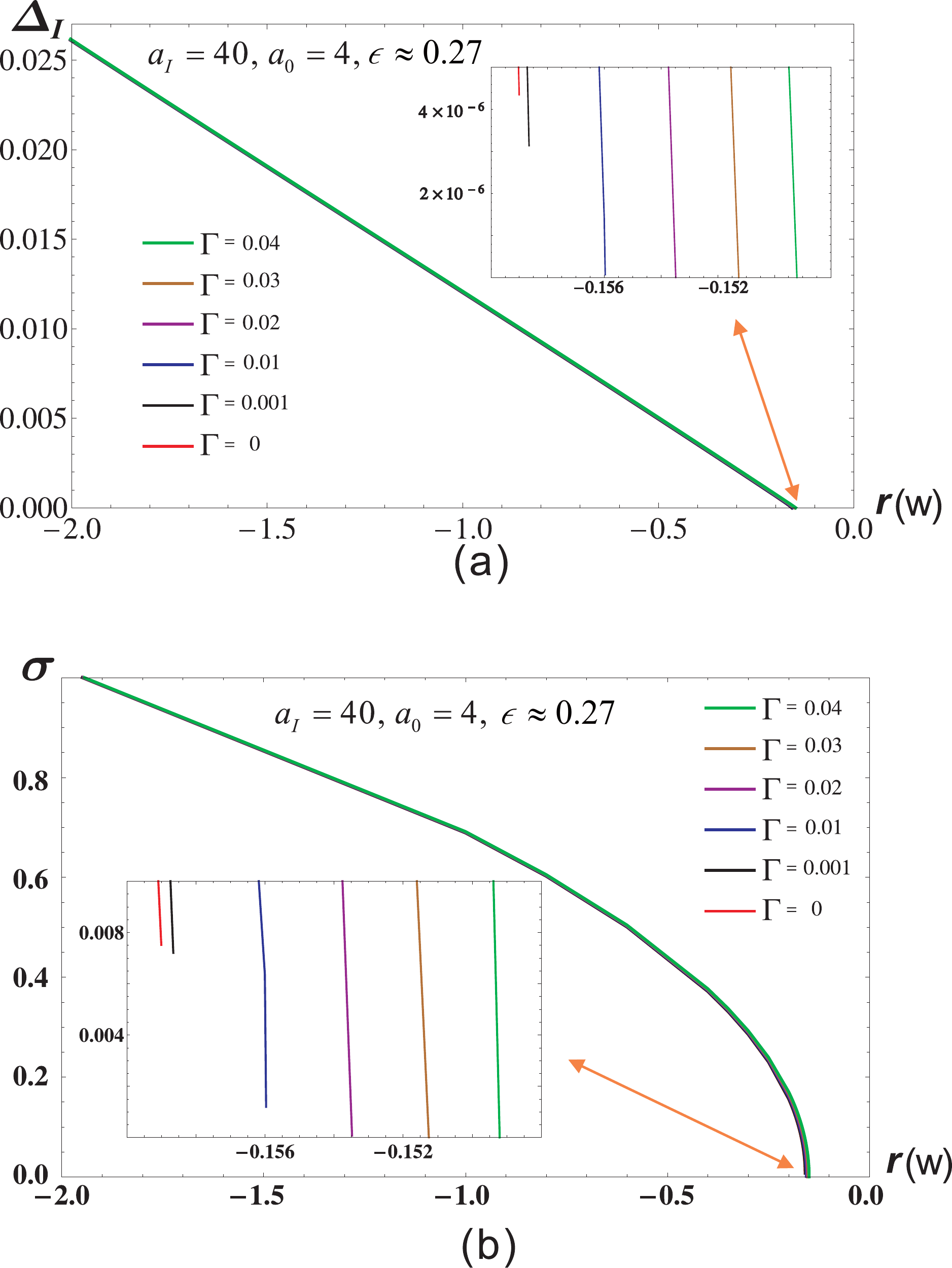}
\end{center}
\caption{The evolution of the Ising order parameter $\Delta_I$ (a) and the collinear AF order parameter $\sigma$ (b)
$vs.$ the control parameter
at different damping rates ($\Gamma = \gamma/(c^{1/2}\Lambda_c)$) at a relatively large anisotropy
$\epsilon \approx 0.27$,
with fixed values of the normalized interactions $a_I$ and $a_0$.
Each order parameter is normalized so that its value deep in the ordered phase is $1$.
The transition is very weakly first order,
with jumps in the order parameters (insets) that are very small and decrease with damping:
already for relatively small damping rate,
the jump is on the order of $10^{-6}$ (and $10^{-3}$) for the nematic (and AF) order parameter.
}
\label{fig:combine1}
\end{figure}

To study the transition more quantitatively, we have solved the large-$N$ equations numerically.
Fig.~\ref{fig:combine1} shows how the Ising and magnetic order parameters change when tuning $w$,
where, for comparison, we assume $r$ can still be tuned even at $\gamma = 0$.
The jump of the order parameters is seen to be very small, even for the case of a relatively large anisotropy:
of ellipticity $\epsilon \approx 0.27$.

As explained in SM (and verified numerically: compare Fig.~\ref{fig:combine1}
and Fig.~S2 for a case of extreme anisotropy with $\epsilon \approx 0.025$),
the order-parameter jump decreases with decreasing anisotropy ({\it i.e.}, increasing ellipticity $\epsilon$).
Experiments in the iron arsenides
observe an ellipticity of $\epsilon \approx 0.7$ \cite{pallab,Diallo2010}, {\it i.e.}
an anisotropy weaker than that shown in Fig.~\ref{fig:combine1}.
We then expect even smaller jumps of the order parameters across the quantum phase transition.

\paragraph*{Effect of the third-dimensional coupling.---}
Iron pnictides have a finite N\'eel temperature, which results from an interlayer exchange coupling.
In order to understand the role of this coupling on the quantum phase transition,
we have studied the effective field theory in three-dimensional space. The details of the model are described
 in the Supplementary Material \cite{supplemental}, and the results for the case
with the spin-wave velocity on the third dimension being equal to the in-plane velocity at $v=0$ are shown
in Figs.~S3,S4. The AF and Ising transitions are still concurrent, and become genuinely continuous.
Again, this is consistent with the RG considerations: given that the effective dimensionality in this case is $d+z=5$,
the quartic coupling $-u_I$ becomes irrelevant w.r.t. the underlying QCP and
will therefore not destabilize the continuous nature of the transition.

In the more general case, with a varying third-dimensional coupling, it is more difficult to solve the large-$N$ equations.
 However, the RG considerations imply that
 turning on the interlayer coupling from the purely 2D limit will further suppress the jump in the order parameters.

\paragraph*{Discussion.---}
Our results imply that the model for the isoelectronically doped iron pnictides yields
quantum phase transitions of the AF and Ising-nematic orders that are concurrent,
and essentially second order. In other words, while in two-dimensions
the transition is eventually
first-order, the jumps of the order parameters are small enough to allow a large dynamical range for quantum criticality;
the smallness of the jumps is ultimately traced
to the marginal nature of the relevant coupling in the effective field theory. In three dimensions,
the transition is continuous. Our conclusion
reconciles the recent observations of quantum criticality in the normal states of P-doped
BaFe$_2$As$_2$ \cite{analytis,analytis2} on the one hand, and the neutron-scattering determination of the weakly
first order nature of the
quantum transition \cite{DHu}.

In addition, the extremely small jump of the order parameters across the quantum phase transition in the two-dimensional case
is also important for understanding other experimental observations.
It implies that quantum criticality occurs over a wide dynamical range,
with two-dimensional character.
The logarithmic divergence of the effective mass expected from such
quantum critical fluctuations
\cite{dai,abrahams} has received considerable experimental support in the
P-doped BaFe$_2$As$_2$.
It fits well the P-doping dependence of the effective mass as extracted
from the de Haas-van Alphen (dHvA) measurements \cite{shishido},
as well as that of the square root of the
$T^2$-coefficient of the electrical resistivity \cite{analytis}.
Finally,
initial dynamical evidence
for quantum critical fluctuations in the antiferromagnetic and Ising-nematic channels
has
come from inelastic neutron scattering measurements in the
electron-doped BaFe$_2$As$_2$ detwinned by uniaxial strain \cite{lu,song2015};
it would be very instructive to explore similar effects in the P-doped BaFe$_2$As$_2$.

\paragraph*{Conclusion.---} We studied zero-temperature magnetic and
Ising transitions in a model
for isoelectronically tuned iron pnictides using a large-N approach. We demonstrated that the two transitions
are concurrent at zero temperature. We also showed that the transition
in the presence of damping are essentially continuous; jumps in the order parameters
are extremely small,
and are further suppressed by an inter-plane coupling.
Our results imply the occurrence of quantum criticality in the isoelectronically
doped iron pnictides,
and reconcile several seemingly contradictory
experimental observations in the P-doped iron arsenides.

\section{Acknowledgement}
We thank J. G. Analytis, P. Dai, W. Ding, A. H. Nevidomskyy, J. H. Pixley and Z. Wang for useful discussions.
The work has been supported in part by the NSF Grant No.\ DMR-1309531 and
the Robert A.\ Welch Foundation Grant No.\ C-1411 (at Rice, J.W. and Q.S.)
and by  the AFOSR Grant No.\ FA9550-14-1-0168 (at UCSD, J.W.).
Q.S.\  acknowledges the support of the Alexander von Humboldt Foundation,
and the hospitality of the the Karlsruhe Institute
of Technology and the Institute of Physics of Chinese Academy of Sciences.

\newpage
\onecolumngrid
\setcounter{figure}{0}
\makeatletter
\renewcommand{\thefigure}{S\@arabic\c@figure}
\setcounter{equation}{0} \makeatletter
\renewcommand \theequation{S\@arabic\c@equation}
\section*{\Large Supplemental Material -- Magnetic and Ising quantum phase transitions in a model for isoelectronically tuned iron pnictides}
{\hfill Jinda Wu, Qimiao Si, and Elihu Abrahams \hfill}

\section{E\MakeLowercase{ffective} A\MakeLowercase{ction}  \MakeLowercase{at} L\MakeLowercase{arge} $N$}
The action, from the main text, Eqs.~(2,3) is $S(\{m\}) = S_2  + S_4$, where
\bea
 S_2  &=& \sum\limits_{\overset{\lower0.5em\hbox{$\smash{\scriptscriptstyle\rightharpoonup}$}} {q} ,\omega _l }
 {\left\{ {\chi _{0,\overset{\lower0.5em\hbox{$\smash{\scriptscriptstyle\rightharpoonup}$}} {q} ,\omega _l }^{ - 1}
 \left( {\left| {\overset{\lower0.5em\hbox{$\smash{\scriptscriptstyle\rightharpoonup}$}}
 {m} _{A,\overset{\lower0.5em\hbox{$\smash{\scriptscriptstyle\rightharpoonup}$}} {q} ,\omega _l } } \right|^2
 + \left| {\overset{\lower0.5em\hbox{$\smash{\scriptscriptstyle\rightharpoonup}$}} {m} _{B,
 \overset{\lower0.5em\hbox{$\smash{\scriptscriptstyle\rightharpoonup}$}} {q} ,\omega _l } } \right|^2 } \right)
 + 2v\left( {q_x^2  - q_y^2 } \right)\overset{\lower0.5em\hbox{$\smash{\scriptscriptstyle\rightharpoonup}$}} {m} _{A,\overset{\lower0.5em
 \hbox{$\smash{\scriptscriptstyle\rightharpoonup}$}} {q} ,\omega _l }
 \cdot \overset{\lower0.5em\hbox{$\smash{\scriptscriptstyle\rightharpoonup}$}} {m} _{B, -
 \overset{\lower0.5em\hbox{$\smash{\scriptscriptstyle\rightharpoonup}$}} {q} , - \omega _l } } \right\}}   \\
  S_4  &=& \int_0^\beta  {d\tau \int {d^2 \overset{\lower0.5em\hbox{$\smash{\scriptscriptstyle\rightharpoonup}$}}
  {r} \left\{ {u_1 \left[ {\left( {\overset{\lower0.5em\hbox{$\smash{\scriptscriptstyle\rightharpoonup}$}} {m} _A^2 } \right)^2
  + \left( {\overset{\lower0.5em\hbox{$\smash{\scriptscriptstyle\rightharpoonup}$}} {m} _B^2 } \right)^2 }
  \right] + u_2\; \overset{\lower0.5em\hbox{$\smash{\scriptscriptstyle\rightharpoonup}$}}
  {m} _A^2 \overset{\lower0.5em\hbox{$\smash{\scriptscriptstyle\rightharpoonup}$}} {m} _B^2  - u_I
  \left( {\overset{\lower0.5em\hbox{$\smash{\scriptscriptstyle\rightharpoonup}$}} {m} _A
  \cdot \overset{\lower0.5em\hbox{$\smash{\scriptscriptstyle\rightharpoonup}$}} {m} _B } \right)^2 } \right\}} }
\eea
where the ${\vec m}_{A/B}$ are $O(N)$ vector fields of the $A/B$ sublattices in either momentum and Matsubara frequency space ($S_2$) or real space and imaginary time ($S_4$) and $\chi _0^{-1}  = r + \omega _l^2  + c~q^2  + \gamma \left| {\omega _l } \right|$ with $r =r_0 + w A_Q$. The quartic couplings have been rescaled by a factor $1/N$ and in the functional integral for the free energy  ${\cal F} = -\ln \int{\cal D}\{m\}\exp(-S(\{m\})$, they can be decoupled as follows:
\be
  e^{(u_I/N) \int {dx\left( {\overset{\lower0.5em\hbox{$\smash{\scriptscriptstyle\rightharpoonup}$}} {m} _A
  \cdot \overset{\lower0.5em\hbox{$\smash{\scriptscriptstyle\rightharpoonup}$}} {m} _B } \right)^2 } }
  = L_1 \int {D\Delta _I e^{\int {dx\left( { - \frac{{N\Delta _I^2 }}
{{u_I }} - 2\Delta _I \overset{\lower0.5em\hbox{$\smash{\scriptscriptstyle\rightharpoonup}$}} {m} _A
 \cdot \overset{\lower0.5em\hbox{$\smash{\scriptscriptstyle\rightharpoonup}$}} {m} _B } \right)} } }  \label{hsdecoupling1}
\ee
and
\bea
&  e^{ - r\sum\limits_{\overset{\lower0.5em\hbox{$\smash{\scriptscriptstyle\rightharpoonup}$}} {q} ,\omega _l }
 {\left( {\left| {\overset{\lower0.5em\hbox{$\smash{\scriptscriptstyle\rightharpoonup}$}}
  {m} _{A,\overset{\lower0.5em\hbox{$\smash{\scriptscriptstyle\rightharpoonup}$}} {q} ,\omega _l } } \right|^2
  + \left| {\overset{\lower0.5em\hbox{$\smash{\scriptscriptstyle\rightharpoonup}$}}
  {m} _{B,\overset{\lower0.5em\hbox{$\smash{\scriptscriptstyle\rightharpoonup}$}} {q} ,\omega _l } } \right|^2 } \right) - }
  \int {dx\left\{ {(u_1/N) \left[ {\left( {\overset{\lower0.5em\hbox{$\smash{\scriptscriptstyle\rightharpoonup}$}}
  {m} _A^2 } \right)^2  + \left( {\overset{\lower0.5em\hbox{$\smash{\scriptscriptstyle\rightharpoonup}$}} {m} _B^2 } \right)^2 } \right]
  + (u_2/N) \overset{\lower0.5em\hbox{$\smash{\scriptscriptstyle\rightharpoonup}$}}
  {m} _A^2 \overset{\lower0.5em\hbox{$\smash{\scriptscriptstyle\rightharpoonup}$}} {m} _B^2 } \right\}} }  \label{hsdecoupling2}  \\
&   = L_2 \int {D\lambda _A D\lambda _B e^{ - i\lambda _A m_A^2  - i\lambda _B m_B^2 } e^{\frac{1}
{2}\int {dx\left( {i\lambda _A  - r,i\lambda _B  - r} \right)\frac{{Nu_1 }}
{{4u_1^2  - u_2^2 }}\left( {\begin{array}{*{20}c}
   2 & { - u_2 /u_1 }  \\
   { - u_2 /u_1 } & 2  \\
 \end{array} } \right)\left( \begin{subarray}{l}
  i\lambda _A  - r \\
  i\lambda _B  - r
\end{subarray}  \right)} } }  \nonumber
\eea
with the normalized factors
\be
L_1  = \prod\limits_x {\sqrt {\frac{{u_I /N}}
{\pi }} } ,\;L_2  = \prod\limits_x {\sqrt {\frac{{\left( {4u_1^2  - u_2^2 } \right)/N^2 }}
{{4\pi ^2 }}} },
\ee
where $x = (\tau, \overset{\lower0.5em\hbox{$\smash{\scriptscriptstyle\rightharpoonup}$}} {r})$ with $\int {dx}
\equiv \int_0^\beta  {d\tau \int {d^2 \overset{\lower0.5em\hbox{$\smash{\scriptscriptstyle\rightharpoonup}$}} {r} } }$.
Eq.~(\ref{hsdecoupling2}), for the case with positive quartic couplings, corresponds to the standard
Hubbard-Stratonovich transformation.
Eq.~(\ref{hsdecoupling1}) describes the case of a negative quartic coupling and a regularization
is needed \cite{Negele1998}.
The LHS of Eq.~(\ref{hsdecoupling1}) will diverge after functional integrations over the ${\vec m}$-fields, which indicates
 that the functional integrals over ${\vec m}_{A/B}$ cannot be interchanged with the functional integral over the field $\Delta_I$ in the RHS of Eq.~(\ref{hsdecoupling1}). However, since
in our case $u_1$ is larger than $u_I$, when we combine the functional integrals over the LHS's
of Eqs.~(\ref{hsdecoupling1},\ref{hsdecoupling2}), the total partition function is regular.
(Another way of seeing this is that  the solutions to the saddle-point equations are bounded.)
 As a result, when we deal with the decoupling over the quartic terms simultaneously, the functional integrals over the fields $\overset{\lower0.5em\hbox{$\smash{\scriptscriptstyle\rightharpoonup}$}}
 {m} _{A/B}$
can be interchanged with those over the
 conjugate fields of $\Delta_I$, $\lambda_A$ and $\lambda_B$.
 Hence, after decoupling the quartic terms we can first integrate over the fluctuations in the fields
 $\overset{\lower0.5em\hbox{$\smash{\scriptscriptstyle\rightharpoonup}$}} {m} _{A}$
 and $\overset{\lower0.5em\hbox{$\smash{\scriptscriptstyle\rightharpoonup}$}} {m} _{B}$,
 leading to the standard procedure of a large-$N$ approach.
 To the leading order in $1/N$, we may, as usual, keep only the zeroth mode ($\omega = 0$, $k=0$) of $i \lambda_{A/B}$
and  $\Delta_I$.
We then integrate over the $(N-1)$ component fluctuation fields
$\overset{\lower0.5em\hbox{$\smash{\scriptscriptstyle\rightharpoonup}$}} {\pi } _{A/B}$
 in $\overset{\lower0.5em\hbox{$\smash{\scriptscriptstyle\rightharpoonup}$}} {m} _{A/B}
  = \left( {\sqrt N \sigma _{A/B} ,\overset{\lower0.5em\hbox{$\smash{\scriptscriptstyle\rightharpoonup}$}} {\pi } _{A/B} } \right)$,
   leaving us with an effective action as a function
of  $\lambda_{A/B}, \sigma_{A/B}$ and $\Delta_I$. Because the sublattices $A$ and $B$ are symmetric,
we have $i\lambda _A  = \left\langle {m_A^2 } \right\rangle  = i\lambda _B  = \left\langle {m_B^2 } \right\rangle = m^2 $,
 and $\sigma _A  =  \pm \sigma _B  = \sigma$. Thus to the order of $O(1/N)$ we get the effective free energy:
\be
{\cal F} = \frac{\Delta _I^2}
{u_I} - \frac{(m^2  - r)^2}
{2u_1  + u_2} + (m^2  \pm \Delta _I )\sigma ^2  + g(m^2,\Delta_I)
\label{freea}
\ee
with
\begin{align}
  g(m^2,\Delta_I) &= \frac{1}{2\beta V}\sum_{{\vec q},l} \ln\big\{(D_{0,{\vec q},l}^{-1}  + m^2)^2 \notag\\
   &-[v(q_x^2  - q_y^2) + \Delta _I]^2\big\}, \label{ironpnictides:g}
\end{align}
where $D_{0,\overset{\lower0.5em\hbox{$\smash{\scriptscriptstyle\rightharpoonup}$}} {q} ,\omega _l }^{ - 1}
= \chi _{0,\overset{\lower0.5em\hbox{$\smash{\scriptscriptstyle\rightharpoonup}$}} {q} ,\omega _l }^{ - 1}  - r$, and we take
$+$ when $\sigma_A = \sigma_B =\sigma$, and $-$ when $\sigma_A = -\sigma_B = \sigma$ in the expression
of $ (m^2 \pm \Delta_I) \sigma^2 $.

\section{S\MakeLowercase{addle} P\MakeLowercase{oint} E\MakeLowercase{quations} \MakeLowercase{and}
S\MakeLowercase{ome} G\MakeLowercase{eneral} C\MakeLowercase{onclusions} \MakeLowercase{to} \MakeLowercase{the}
O\MakeLowercase{rder} \MakeLowercase{of} $O(1/N)$}
From Eq.~(\ref{freea}) we have variational equations w.r.t $\sigma$, $m^2$ and $\Delta_I$,
\be
\frac{{\partial {\cal F}}}{{\partial \sigma  }} = \frac{{\partial {\cal F}}}{{\partial \Delta _I }} = \frac{{\partial {\cal F}}}{{\partial m^2 }} =0
\ee
After re-arranging these equations we have (for convenience here we choose the branch $\sigma_A = \sigma_B = \sigma$)
\bea
&\left( { m^2 + \Delta_I } \right)\sigma  = 0 \label{ironpnictides:v1} \\
\frac{{\Delta _I }}{{u_I }} &= \frac{{m^2  - r}}{{2u_1  + u_2 }} - 2 \sigma ^2
- \frac{1}{{2\beta V}}\sum\limits_{\mathord{\buildrel{\lower3pt\hbox{$\scriptscriptstyle\rightharpoonup$}}
\over q} ,i\omega _l } {\frac{1}{{D_{0,\mathord{\buildrel{\lower3pt\hbox{$\scriptscriptstyle\rightharpoonup$}}
\over q} ,i\omega _l }^{ - 1}  + v\left( {q_x^2  - q_y^2 } \right) + m^2  + \Delta _I }}} \label{ironpnictides:v4}  \\
 \frac{{\Delta _I }}{{u_I }} &=  - \frac{{m^2  - r}}{{2u_1  + u_2 }} + \frac{1}{{2\beta V}}
 \sum\limits_{\mathord{\buildrel{\lower3pt\hbox{$\scriptscriptstyle\rightharpoonup$}}
\over q} ,i\omega _l } {\frac{1}{{D_{0,\mathord{\buildrel{\lower3pt\hbox{$\scriptscriptstyle\rightharpoonup$}}
\over q} ,i\omega _l }^{ - 1}  - v\left( {q_x^2  - q_y^2 } \right) + m^2  - \Delta _I }}}.  \label{ironpnictides:v5}
\eea
Eqs.~(\ref{ironpnictides:v4}, \ref{ironpnictides:v5}) imply that, for the branch $\sigma_A = \sigma_B = \sigma$, $\Delta_I \leq 0$.
When $\Delta_I =0$, after summing over Eq.~(\ref{ironpnictides:v4}) and Eq.~(\ref{ironpnictides:v5}) we immediately
have $\sigma = 0$. In other words the vanishing of $\Delta_I $ can not happen before $\sigma$ vanishes.

On the other hand when $\sigma = 0$, Eq.~(\ref{ironpnictides:v4}) and Eq.~(\ref{ironpnictides:v5}) merge to one equation. After doing analytic continuation, then setting $T =0$, this combined equation becomes,
\be
\frac{{2\Delta _I }}
{{u_I }} = \left( {\frac{1}
{{2\pi }}} \right)^3 \int_{ - {\Lambda_f} }^{\Lambda_f}  {d^2 q\int_0^\infty  {d\omega \left[ {\frac{{\gamma \omega }}
{{\left( {\omega ^2  - c_1^2 } \right)^2  + \gamma ^2 \omega ^2 }} - \frac{{\gamma \omega }}
{{\left( {\omega ^2  - c_0^2 } \right)^2  + \gamma ^2 \omega ^2 }}} \right]} } \label{ironpnictides:sum}
\ee
where $\Lambda_f$ is the Fermi wave vector, and
\bea
c_0^2  = \left( {c + v} \right)q_x^2  + \left( {c - v} \right)q_y^2  + m^2  + \Delta _I  \\
c_1^2  = \left( {c - v} \right)q_x^2  + \left( {c + v} \right)q_y^2  + m^2  - \Delta _I.
 \eea
After the integrations on the right hand side (RHS) of Eq.~(\ref{ironpnictides:sum}), it
becomes (see the next section for the detailed calculations)
\bea
\frac{{2\Delta _I }}
{{u_I }} =  \frac{1}
{{16\pi ^2 \sqrt {c^2  - v^2 } }}\left\{ {\gamma \ln \frac{{m^2  - \Delta _I }}
{{m^2  + \Delta _I }} + i\sqrt {4\left( {m^2  + \Delta _I } \right) - \gamma ^2 } \ln \frac{{\gamma  - i\sqrt {4\left( {m^2  + \Delta _I }
\right) - \gamma ^2 } }}
{{\gamma  + i\sqrt {4\left( {m^2  + \Delta _I } \right) - \gamma ^2 } }}} \right. \nonumber \\
  \;\;\;\;\;\;\;\;\;\;\;\;\;\;\;\;\;\;\;\;\;\;\;\;\;\;\;\;\;\;\;\;\;\;\;\;\;\left. { - i\sqrt {4\left( {m^2  - \Delta _I } \right) - \gamma ^2 } \ln \frac{{\gamma
  - i\sqrt {4\left( {m^2  - \Delta _I } \right) - \gamma ^2 } }}
{{\gamma  + i\sqrt {4\left( {m^2  - \Delta _I } \right) - \gamma ^2 } }}} \right\}   \label{ironpnictides:sum2}
\eea
The solution of Eq.~(\ref{ironpnictides:sum2}) is easiest to see in the limit of $m^2/\gamma^2 \ll 1$,
where $\Delta_I = 0$ is the only solution that is
consistent with our particular limit $\sigma=0$.
This is also valid in the other limit but it is more technically involved to demonstrate.
Based on these asymptotic results, we expect that, at zero temperature and to  order $O(1/N)$, vanishing of the magnetic order ($\sigma=0$)
implies that the Ising order also vanishes.  This conclusion is also numerically confirmed.

\section{C\MakeLowercase{alculation} \MakeLowercase{of} S\MakeLowercase{ummations} \MakeLowercase{in}
E\MakeLowercase{q.~}(\ref{ironpnictides:v4},\ref{ironpnictides:v5})}
For nonvanishing magnetic order ($\sigma\neq 0$), we need to evaluate the sums in Eqs.~(\ref{ironpnictides:v4}, \ref{ironpnictides:v5}),as follows:
\bea
  \frac{1}
{{2\beta V}}\sum\limits_{\overset{\lower0.5em\hbox{$\smash{\scriptscriptstyle\rightharpoonup}$}} {q} ,i\omega _l } {\frac{1}
{{D_{0,\overset{\lower0.5em\hbox{$\smash{\scriptscriptstyle\rightharpoonup}$}} {q} ,i\omega _l }^{ - 1}
+ v\left( {q_x^2  - q_y^2 } \right) + m^2  + \Delta _I }}}  = \frac{1}
{{\left( {2\pi } \right)^3 }}\int_{ - \Lambda _f }^{\Lambda _f } {d^2 q\int_0^{\Gamma _0 } {d\omega \coth \frac{\omega }
{{2T}}\frac{{\gamma \omega }}
{{\left( {\omega ^2  - c_0^2 } \right)^2  + \gamma ^2 \omega ^2 }}} }  \hfill \label{ironpnictides:sum01} \\
  \frac{1}
{{2\beta V}}\sum\limits_{\overset{\lower0.5em\hbox{$\smash{\scriptscriptstyle\rightharpoonup}$}} {q} ,i\omega _l } {\frac{1}
{{D_{0,\overset{\lower0.5em\hbox{$\smash{\scriptscriptstyle\rightharpoonup}$}} {q} ,i\omega _l }^{ - 1}  + v\left( {q_x^2  - q_y^2 } \right)
+ m^2  - \Delta _I }}}  = \frac{1}
{{\left( {2\pi } \right)^3 }}\int_{ - \Lambda _f }^{\Lambda _f } {d^2 q\int_0^{\Gamma _0 } {d\omega \coth \frac{\omega }
{{2T}}\frac{{\gamma \omega }}
{{\left( {\omega ^2  - c_1^2 } \right)^2  + \gamma ^2 \omega ^2 }}} }  \hfill \label{ironpnictides:sum02}
\eea
At $T = 0$ K, from Eq.~(\ref{ironpnictides:sum01}) we have
\bea
&&  \left. {\frac{1}
{{\left( {2\pi } \right)^3 }}\int_{ - \Lambda _f }^{\Lambda _f } {d^2 q\int_0^{\Gamma _0 } {d\omega \coth \frac{\omega }
{{2T}}\frac{{\gamma \omega }}
{{\left( {\omega ^2  - c_0^2 } \right)^2  + \gamma ^2 \omega ^2 }}} } } \right|_{T = 0}  = \frac{1}
{{\left( {2\pi } \right)^3 }}\int_{ - \Lambda _f }^{\Lambda _f } {d^2 q\int_0^{\Gamma _0 } {d\omega \frac{{\gamma \omega }}
{{\left( {\omega ^2  - c_0^2 } \right)^2  + \gamma ^2 \omega ^2 }}} }  \\
&\approx& \frac{1}
{{\left( {2\pi } \right)^3 }}\int_{ - \Lambda _f }^{\Lambda _f } {d^2 q\int_0^\infty  {d\omega \frac{{\gamma \omega }}
{{\left( {\omega ^2  - c_0^2 } \right)^2  + \gamma ^2 \omega ^2 }}} }  = \frac{1}
{{\left( {2\pi } \right)^3 }}\int_{ - \Lambda _f }^{\Lambda _f } {d^2 q\frac{i}
{{\sqrt {4c_0^2  - \gamma ^2 } }}\ln \frac{{\gamma  - i\sqrt {4c_0^2  - \gamma ^2 } }}
{{\gamma  + i\sqrt {4c_0^2  - \gamma ^2 } }}} \\
& \approx& \frac{{\gamma \ln \frac{{4(m^2  + \Delta _I )}}
{{4c\Lambda _c^2  + \gamma ^2 }} + i2\sqrt {c\Lambda _c } \ln \frac{{\gamma  - i2\sqrt c \Lambda _c }}
{{\gamma  + i2\sqrt c \Lambda _c }} - i\sqrt {4(m^2  + \Delta _I ) - \gamma ^2 } \ln \frac{{\gamma  - i\sqrt {4(m^2  + \Delta _I ) - \gamma ^2 } }}
{{\gamma  + i\sqrt {4(m^2  + \Delta _I ) - \gamma ^2 } }}}}
{{16\pi ^2 \sqrt {c^2  - v^2 } }}
  = G_1  \label{G1}
\eea
where in the last approximation we have used the approximate condition $\sqrt{c-v} \Lambda_f \gg \sqrt{c} \Lambda_c$,
where the anisotropy is not extremely strong and $\Lambda_c$ is the low-energy cut-off wave vector for spin excitations.
This condition will also be applied for all following calculations. Similarly, the integration in Eq.~(\ref{ironpnictides:sum02})
can be calculated as follows,
\bea
&&   \left. {\frac{1}
{{\left( {2\pi } \right)^3 }}\int_{ - \Lambda _f }^{\Lambda _f } {d^2 q\int_0^{\Gamma _0 } {d\omega \coth \frac{\omega }
{{2T}}\frac{{\gamma \omega }}
{{\left( {\omega ^2  - c_1^2 } \right)^2  + \gamma ^2 \omega ^2 }}} } } \right|_{T = 0}  = \frac{1}
{{\left( {2\pi } \right)^3 }}\int_{ - \Lambda _f }^{\Lambda _f } {d^2 q\int_0^{\Gamma _0 } {d\omega \frac{{\gamma \omega }}
{{\left( {\omega ^2  - c_1^2 } \right)^2  + \gamma ^2 \omega ^2 }}} }   \\
&\approx& \frac{1}
{{\left( {2\pi } \right)^3 }}\int_{ - \Lambda _f }^{\Lambda _f } {d^2 q\int_0^\infty  {d\omega \frac{{\gamma \omega }}
{{\left( {\omega ^2  - c_1^2 } \right)^2  + \gamma ^2 \omega ^2 }}} }  = \frac{1}
{{\left( {2\pi } \right)^3 }}\int_{ - \Lambda _f }^{\Lambda _f } {d^2 q\frac{i}
{{\sqrt {4c_1^2  - \gamma ^2 } }}\ln \frac{{\gamma  - i\sqrt {4c_1^2  - \gamma ^2 } }}
{{\gamma  + i\sqrt {4c_1^2  - \gamma ^2 } }}}  \\
&=& \frac{{\gamma \ln \frac{{4(m^2  - \Delta _I )}}
{{4c\Lambda _c^2  + \gamma ^2 }} + i2\sqrt {c\Lambda _c } \ln \frac{{\gamma  - i2\sqrt c \Lambda _c }}
{{\gamma  + i2\sqrt c \Lambda _c }} - i\sqrt {4(m^2  - \Delta _I ) - \gamma ^2 } \ln \frac{{\gamma  - i\sqrt {4(m^2
- \Delta _I ) - \gamma ^2 } }}
{{\gamma  + i\sqrt {4(m^2  - \Delta _I ) - \gamma ^2 } }}}}
{{16\pi ^2 \sqrt {c^2  - v^2 } }}
= G_2  \label{G2}
\eea

Using Eqs.~(\ref{G1}, \ref{G2}), after some integrals, we can get an analytical expression for the free energy
as a function of $\Delta_I, m^2,  \sigma$. The central task here is to get a closed form for Eq.~(\ref{ironpnictides:g}).
We can tackle the summation as follows,
\be
\frac{{\partial g}}{{\partial \Delta _I }} = G_1  - G_2  \equiv g'_{\Delta _I } ;\;\;\;\frac{{\partial g}}{{\partial m^2 }} = G_1
+ G_2  \equiv g'_{^{m^2 } }
\ee
Then we have
\be
g(\Delta _I ,m^2 ) = \int_0^{\Delta _I } {g'_{\Delta _I } (x,m^2 )dx}  + \int_0^{m^2 } {g'_{^{m^2 } } (0,y)dy}
\ee
After finishing integrations in the above equation, we can get a closed form of $g(\Delta_I, m^2)$.
Substituting the closed form back into Eq.~(\ref{freea}),
we arrive at the expression for the full free energy given by
\be
\begin{gathered}
  {\cal F} = \frac{{\Delta _I^2 }}
{{u_I }} - \frac{{\left( {m^2  - r} \right)^2 }}
{{2u_1  + u_2 }} + 2\left( {m^2  - |\Delta _I| } \right)\sigma ^2  + \frac{{\gamma ^3 a_c }}
{{16 c \pi ^2  }}\left\{ {\left( {x - \frac{1}
{6}} \right)\ln x + \left( {y - \frac{1}
{6}} \right)\ln y - \left( {x + y} \right)\left[ {\frac{1}
{3} + \ln \left( {1 + 4\frac{{c\Lambda _c^2 }}
{{\gamma ^2 }}} \right)} \right]} \right. \hfill \\
  \;\;\;\;\;\left. { - \frac{1}
{6}\left( {1 - 4y} \right)^{3/2} \ln \frac{{1 + \left( {1 - 4y} \right)^{1/2} }}
{{1 - \left( {1 - 4y} \right)^{1/2} }} - \frac{1}
{6}\left( {1 - 4x} \right)^{3/2} \ln \frac{{1 + \left( {1 - 4x} \right)^{1/2} }}
{{1 - \left( {1 - 4x} \right)^{1/2} }} + \frac{{4\sqrt c \Lambda _c }}
{\gamma }\left( {x + y} \right)\tan ^{ - 1} \frac{{2\sqrt c \Lambda _c }}
{\gamma }} \right\}, \hfill  \\
\end{gathered} \label{freeb}
\ee
where we have introduced the notations $x = (m^2 +\Delta_I)/\gamma^2$ and $y = (m^2 - \Delta_I)/\gamma^2$
with the physical requirement $m^2 \ge |\Delta_I|$,
which guarantees the free energy to be  real.

\section{S\MakeLowercase{addle} P\MakeLowercase{oint} E\MakeLowercase{quations} \MakeLowercase{in}
\MakeLowercase{the} O\MakeLowercase{rdered} R\MakeLowercase{egime}}
From Eqs.~(\ref{ironpnictides:v1},\ref{ironpnictides:v4},\ref{ironpnictides:v5},\ref{G1},\ref{G2}), we arrive at
the following forms of the saddle-point equations in the ordered regime,
\bea
 - \left(  {\frac{1}
{{u_I }} - \frac{1}
{{2u_1  + u_2 }}} \right)\left| {\Delta _I } \right| = \frac{{r(w)}}
{{2u_1  + u_2 }} + \frac{1}
{{16\pi ^2 \sqrt {c^2  - v^2 } }}\left\{ {\gamma \ln \frac{{8\left| {\Delta _I } \right|}}
{{4c\Lambda _c^2  + \gamma ^2 }} + 4\sqrt c \Lambda _c \tan ^{ - 1} \left( {\frac{{2\sqrt c \Lambda _c }}
{\gamma }} \right)} \right. \nonumber \\
  \left. {\;\;\;\;\;\;\;\;\;\;\;\;\;\;\;\;\;\;\;\;\;\;\;\;\;\;\;\;\;\;\;\;\;\;\;\;\;\;\;\;\;\;\;\;\;\;\;\;\;\;\;\;\;\;\;\;\;\;\;\;\;\;\;\;\;\;\;\;\;\; + \sqrt {\gamma ^2
  - 8\left| {\Delta _I } \right|} \ln \frac{{\gamma  + \sqrt {\gamma ^2  - 8\left| {\Delta _I } \right|} }}
{{\gamma  - \sqrt {\gamma ^2  - 8\left| {\Delta _I } \right|} }}} \right\}
\eea
and
\be
2\sigma ^2  =  - \frac{{2\left| {\Delta _I } \right|}}
{{u_I }} + \frac{1}
{{16\pi ^2 \sqrt {c^2  - v^2 } }}\left\{ {\gamma \ln \frac{{2\left| {\Delta _I } \right|}}
{{\gamma ^2 }} + \sqrt {\gamma ^2  - 8\left| {\Delta _I } \right|} \ln \frac{{\gamma  + \sqrt {\gamma ^2  - 8\left| {\Delta _I } \right|} }}
{{\gamma  - \sqrt {\gamma ^2  - 8\left| {\Delta _I } \right|} }}} \right\}
\ee

\begin{figure}[t!]
\begin{center}
\includegraphics[width=13.5cm]{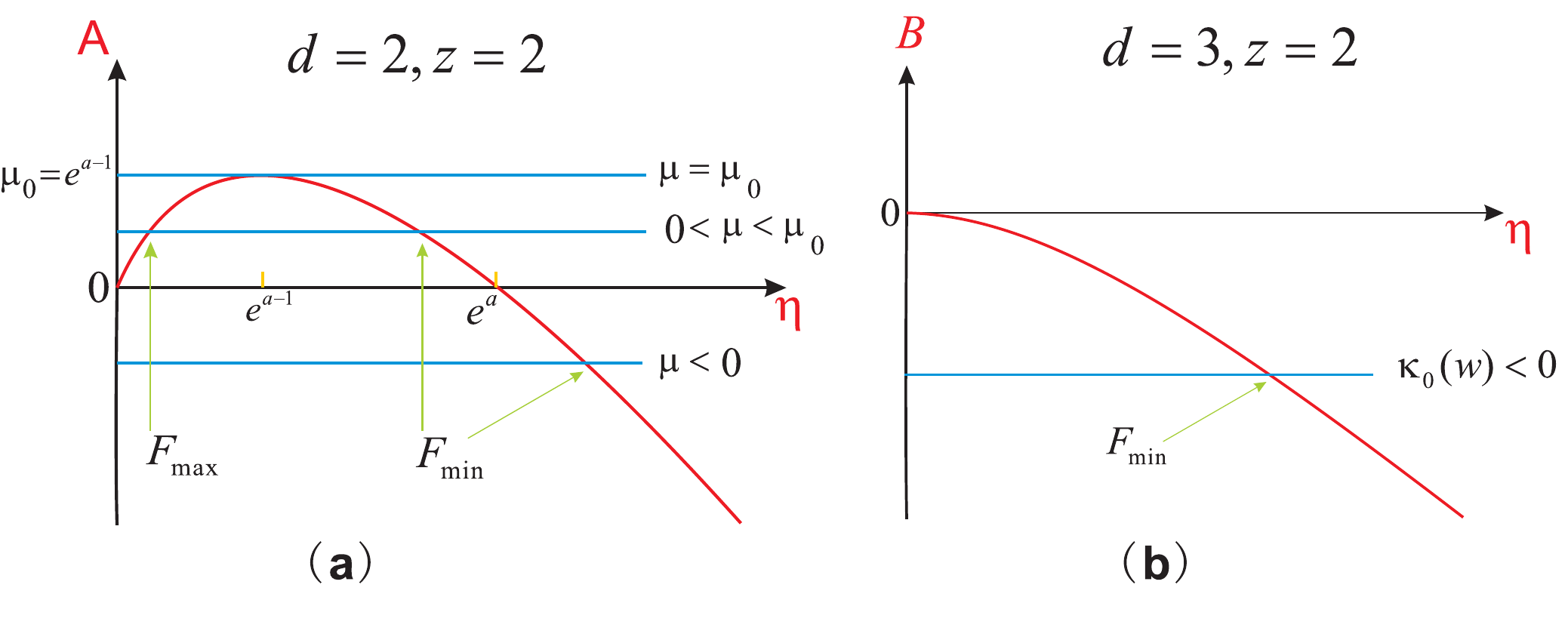}
\end{center}
\caption[Sketchy processes for possible phase transitions in d2z2 and d3z2 systems]{
(a) Illustration of Eq.~(\ref{sm:leadingising});
(b) The counterpart for the generic case of $a_{d3z2} < 0$ in 3D, illustrating
Eq.~(\ref{ironpnictides:d3z2variational}), with $\eta = |\delta|/8.$}
\label{fig:sketchprocesscombine}
\end{figure}

\section{N\MakeLowercase{ature} \MakeLowercase{of} \MakeLowercase{the} M\MakeLowercase{agnetic} \MakeLowercase{and}
 I\MakeLowercase{sing} T\MakeLowercase{ransitions} \MakeLowercase{at} Z\MakeLowercase{ero} T\MakeLowercase{emperature}}
We consider here the concurrent magnetic and Ising transitions at $T=0$.
The RG arguments we outlined in the main text suggest that there will be a jump in the order parameters
across the transitions, but the jump will be smaller as the damping parameter $\gamma$ increases.
To see how damping affects the transition,  we consider the parameter regime
where analytical insights can be gained in our large-$N$ approach.
When $\gamma$ is sufficiently large so that
$x, y \ll 1$ (definitions of $x,y$ are given in the main text), it follows from the closed form of free energy [Eq.~(7) in the main text] that
\be
  \frac{{\cal F}}
{{c^{1/2} \Lambda _c^3 }} =    - a_0 \left( {\frac{{r(w)}}
{{c\Lambda _c^2 }}} \right)^2  + \frac{{\Gamma ^3 a_c }}
{{2\pi ^2 }}\mu (w)m_0^2  + 2\Gamma ^2 \left( {m _0^2  - \left| {\delta _0 } \right|} \right)\sigma _0^2  +  \cdots
 \label{freeapp}
\ee
where in ``$\cdots$'' we temporarily neglect terms at the order of $O[|\delta_0|^2 \ln |\delta_0|]$ and $O[m_0^4 \ln m_0^2]$,
which will be restored when getting Eq.~(\ref{sm:leadingising}). Also $m_0^2  = m^2 /\gamma ^2 = (x+y)/2,\;\delta _0  = \Delta_I /\gamma ^2
= (x-y)/2, \;\sigma _0^2  = \sigma ^2 /\left( {c^{ - 1/2} \Lambda _c } \right)$. In addition $a_c$ and $a_0,\;a_, \; \mu(w), \; \Gamma$
are respectively defined in Eq.(6) and Eq.(14) in the main text.
The $a_c$ here is related to the ellipticity $\epsilon$ by $a_c = (\epsilon +1/\epsilon)/2 \geq 1$;
therefore, a larger $a_c$  means a stronger anisotropy for the system. From Eq.~(\ref{freeapp}), we see that
if $r(w)$ is a large positive number, the minimum of the free energy only occurs at  $\sigma_0 = 0$ and $m_0 = 0$,
then $\Delta_I  = 0$, corresponding to the disordered phase of the system as expected. Eq.~(\ref{freeapp}) shows
that when the system is deep inside the ordered phase with $r_0 < 0$ and $|r_0| \gg 1$, there is no minimum at the origin
since $\mu(w) < 0$. This implies that when we increase $r$ from a large negative value (deep in the ordered phase)
to a certain critical point a phase transition must happen. This can be made clearer when the system stays
in the ordered regime ($\sigma \neq 0$). Here in the limit $ \eta \equiv |\delta_0| \ll 1$,
to order of $(\left| {\Delta _I } \right|/\gamma ^2)^2$, we get
\be
A(\eta)= a \eta - \eta\ln \eta = \mu(w) \label{sm:leadingising}
\ee
with
\bea
\begin{gathered}
  a =  - \frac{{4\pi ^2 \Gamma\left( {a_I  - a_0 } \right)}}
{{a_c }} - \ln 2 - 1/2. \hfill \\
\end{gathered} \label{sm:ironpnictides:a}
\eea
We see that $a < 0$ generally holds, which means the maximum of $A(\eta)$ will be $\mu_0 = e^{a-1}$ at $\eta_0 = e^{a-1}$.
The evolution of the equation is illustrated in Fig.~\ref{fig:sketchprocesscombine}(a). When the system is in the ordered regime,
i.e., $r(w)$ is a large negative number, then $\mu < 0$ and there is a unique global minimum (we focus on the positive branch
of the Ising order parameter). When $r(w)$ increases (via increasing $w$) to the point that $\mu  = 0 $, there is a maximum emerging
at the origin while the Ising order shrinks to $\eta_1 = e^a$. After this, when $r(w)$ is further increased, the maximum emerges at the
origin moves away from the origin with a cusp-type local minimum generated at the origin which can not be covered
by Eqs.~(\ref{ironpnictides:v5},\ref{sm:leadingising}), meanwhile the Ising order shrinks further. When $r(w)$ is further increased
until $\mu  = \mu_0$, the local maximum and local minimum merge as an inflection point, and the free energy as a function
of Ising order will only have a global cusp-type minimum at the origin. Therefore a first order transition happens when
$e^{a-1} < \eta < e^a$, while tuning $w$ to $w_c$ such that $0 < \mu(w_c) < \mu_0$. From Eq.~(\ref{sm:ironpnictides:a}) we can
see larger $\Gamma$ leads to more negative $a$, since the transition happens in the regime of $e^{a-1} < \eta < e^a $,
as a result, the first order transition would be exponentially suppressed when $\Gamma$ becomes larger, implying the transition
would become essentially second order when damping becomes strong, which is consistent with RG predictions.

\begin{figure}[t!]
\begin{center}
\includegraphics[width=16.5cm]{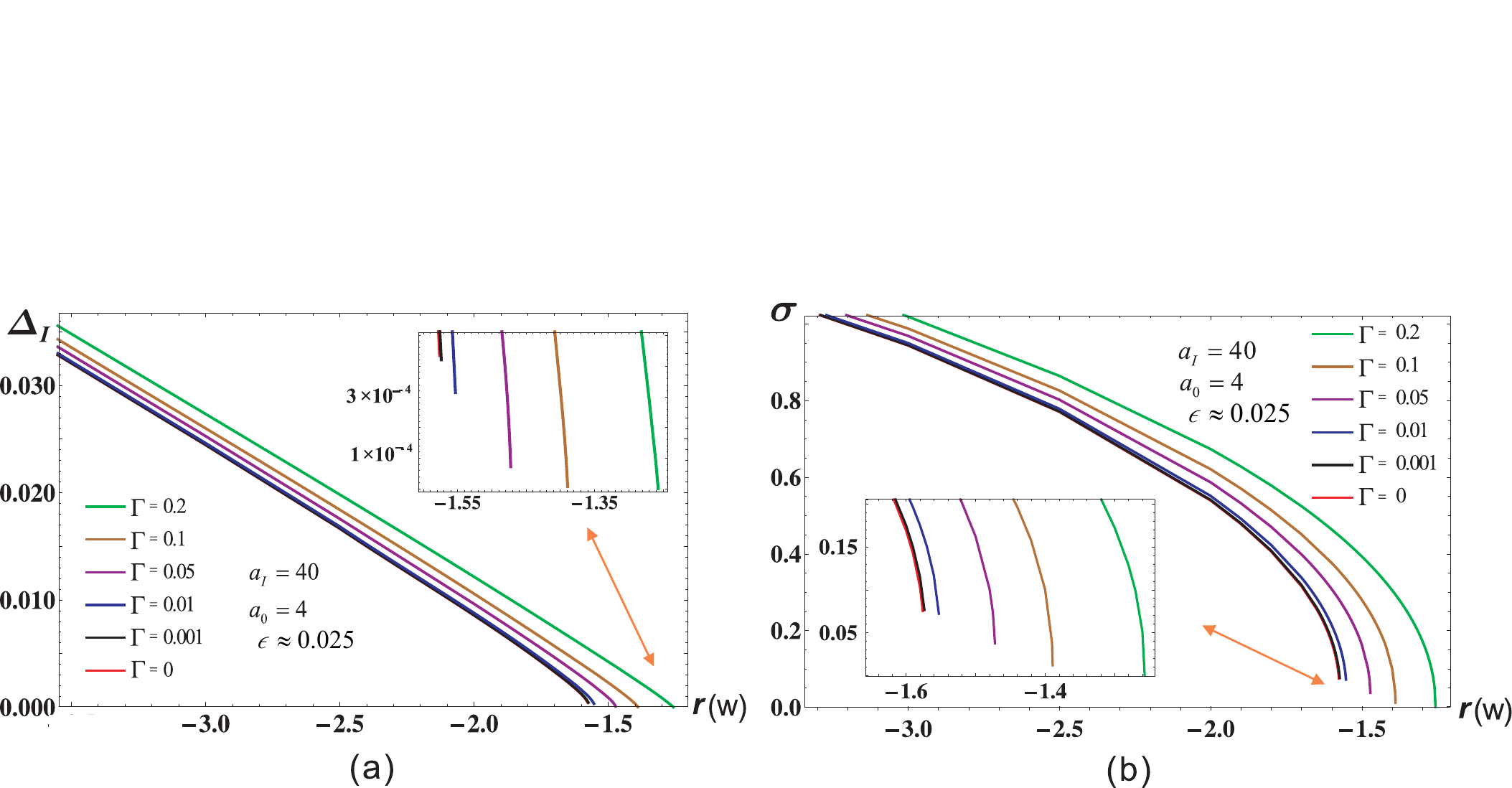}
\end{center}
\caption{The evolution of Ising order (a) and antiferromagnetic order (b) as a function of the control parameter
$r(w)$ at an extremely strong anisotropy $\epsilon \approx 0.025$. The jump of the order parameters becomes larger compared with the case of a moderately strong anisotropy $\epsilon \approx 0.27$ shown in Fig.~2 of the main text.}
\label{fig:d2z2extreme}
\end{figure}

\section{T\MakeLowercase{he} E\MakeLowercase{ffect} \MakeLowercase{of}
E\MakeLowercase{xtreme} A\MakeLowercase{nisotropy}}
When the anisotropy becomes extremely large, the system effectively becomes 1D, and the effective dimensionality
$d+z$ becomes 3; the quartic coupling $-u_I$ will become relevant (as opposed to being marginal) w.r.t.
the underlying O($3$) QCP, and we expect a stronger degree of first-orderness. Indeed, as shown in
Fig.~\ref{fig:d2z2extreme} for an extreme value of anisotropy  $\epsilon=0.025$, the magnetic order parameter jump becomes sizable.

\section{T\MakeLowercase{he} C\MakeLowercase{ase} \MakeLowercase{of} T\MakeLowercase{hree} S\MakeLowercase{patial}
D\MakeLowercase{imensions}}
In this case, we still have the same saddle-point equation Eq.~(\ref{ironpnictides:v5}), but now we take $q^2 = q_x^2
+q_y^2+q_z^2$ in ${\chi _{0,\overset{\lower0.5em\hbox{$\smash{\scriptscriptstyle\rightharpoonup}$}} {q} ,\omega _l }^{ - 1} }
 = r + \omega _l^2  + cq^2  + \gamma \left| {\omega _l } \right|$. Then at zero temperature the summation in Eq.~(\ref{ironpnictides:v5}) can be
calculated as follows (using Eq.~(\ref{ironpnictides:sum02}) and working in the regime of $\Delta_I = -m^2 < 0$).
\bea
&&\frac{1}{{2\beta V}}\sum\limits_{\overset{\lower0.5em\hbox{$\smash{\scriptscriptstyle\rightharpoonup}$}} {q} ,i\omega _l } {\frac{1}
{{D_{0,\overset{\lower0.5em\hbox{$\smash{\scriptscriptstyle\rightharpoonup}$}} {q} ,i\omega _l }^{ - 1}  + v\left( {q_x^2  - q_y^2 } \right)
+ m^2  - \Delta _I }}}  = \frac{1}
{{\left( {2\pi } \right)^4 }}\int_{ - \Lambda _f }^{\Lambda _f } {d^3 q\int_0^{\Gamma _0 } {d\omega \frac{{\gamma \omega }}
{{\left( {\omega ^2  - c_1^2 } \right)^2  + \gamma ^2 \omega ^2 }}} }   \\
   &\approx& \frac{1}
{{8\pi ^3 }}\frac{1}
{{c^{1/2} \sqrt {c^2  - v^2 } }}\int_0^{c\Lambda _c^2 } {dx\frac{{i\sqrt x }}
{{\sqrt {4x + 4\left( {m^2  - \Delta _I } \right) - \gamma ^2 } }}\ln \frac{{\gamma  - i\sqrt {4x + 4\left( {m^2  - \Delta _I } \right) - \gamma ^2 } }}
{{\gamma  + i\sqrt {4x + 4\left( {m^2  - \Delta _I } \right) - \gamma ^2 } }}} \\
   &=& \frac{1}
{{64\pi ^3 }}\frac{\gamma^2}
{{c^{1/2} \sqrt {c^2  - v^2 } }}\int_0^{\Lambda _\gamma  } {dz\frac{{i\sqrt z }}
{{\sqrt {z - \left( {1 + \delta } \right)} }}\ln \frac{{1  - i\sqrt {z - \left( {1 + \delta } \right)} }}
{{1  + i\sqrt {z - \left( {1 + \delta } \right)} }}}  \\
 &=& \frac{1}
{{64\pi ^3 }}\frac{\gamma^2}
{{c^{1/2} \sqrt {c^2  - v^2 } }}\left\{ {\underbrace {2i\int_{\sqrt {\left| \delta  \right| - 1} }^0 {dx\sqrt {x^2  + 1 - \left| \delta  \right|} \ln \frac{{1 - ix}}
{{1 + ix}}} }_I + \underbrace {2i\int_0^{\sqrt {\Lambda _\gamma  } } {dx\sqrt {x^2  + 1 - \left| \delta  \right|} \ln \frac{{1 - ix}}
{{1 + ix}}} }_{II}} \right\}  \label{ironpnictides:d3z2integrals}
\eea
where $\Lambda _\gamma   = 4c\Lambda _c^2 /\gamma ^2 = 4/\Gamma^2 ,\;\;\delta  = 8\Delta _I /\gamma ^2$.
Now let's deal with the two integrals one by one.
\bea
  I &=& 2\int_0^{\left( {1 - \left| \delta  \right|} \right)^{1/2} } {dx\sqrt {1 - \left| \delta  \right| - x^2 } \ln \frac{{1 + x}}
{{1 - x}}}  \\
   &=& 2\left( {1 - \left| \delta  \right|} \right)^{1/2} \sum\limits_{n = 0}^\infty  {\frac{{( - 1)^n }}
{{\left( {1 - \left| \delta  \right|} \right)^n }}\left( \begin{array}{l}
 n \\
 1/2 \\
 \end{array} \right)\int_0^{\left( {1 - \left| \delta  \right|} \right)^{1/2} } {dxx^{2n} \ln \frac{{1 + x}}
{{1 - x}}} }  \\
   &=& 2\left( {1 - \left| \delta  \right|} \right)^{1/2} \sum\limits_{n = 0}^\infty  {\frac{{( - 1)^n }}
{{\left( {1 - \left| \delta  \right|} \right)^n }}\left( \begin{array}{l}
 n \\
 1/2 \\
 \end{array} \right)\left[ {\frac{{\left( {1 - \left| \delta  \right|} \right)^{n + 1/2} }}
{{2n + 1}}\ln \frac{{1 + \left( {1 - \left| \delta  \right|} \right)^{1/2} }}
{{1 - \left( {1 - \left| \delta  \right|} \right)^{1/2} }} - \int_0^{\left( {1 - \left| \delta  \right|} \right)^{1/2} } {\frac{{dxx^{2n + 1} }}
{{2n + 1}}\frac{2}
{{1 - x^2 }}} } \right]}  \\
   &=& 2\frac{\pi }
{4}\left( {1 - \left| \delta  \right|} \right)\ln \frac{{1 + \left( {1 - \left| \delta  \right|} \right)^{1/2} }}
{{1 - \left( {1 - \left| \delta  \right|} \right)^{1/2} }} + 2\left( {1 - \left| \delta  \right|} \right)^{1/2} \sum\limits_{n = 0}^\infty  {\frac{{( - 1)^n }}
{{\left( {1 - \left| \delta  \right|} \right)^n }}\left( \begin{array}{l}
 1/2 \\
 n \\
 \end{array} \right)\left[ { - \int_0^{\left( {1 - \left| \delta  \right|} \right)^{1/2} } {\frac{{dxx^{2n + 1} }}
{{2n + 1}}\frac{2}
{{1 - x^2 }}} } \right]}  \\
   &=& 2\frac{\pi }
{4}\left( {1 - \left| \delta  \right|} \right)\ln \frac{{1 + \left( {1 - \left| \delta  \right|} \right)^{1/2} }}
{{1 - \left( {1 - \left| \delta  \right|} \right)^{1/2} }} + 2\left( {1 - \left| \delta  \right|} \right)^{1/2}
\int_0^{\left( {1 - \left| \delta  \right|} \right)^{1/2} } {dx\frac{{x\sqrt {1 - \frac{{x^2 }}
{{1 - \left| \delta  \right|}}}  + \sqrt {\left| \delta  \right| - 1} \sinh ^{ - 1} \left( {\frac{x}
{{\sqrt {\left| \delta  \right| - 1} }}} \right)}}
{{ - 1 + x^2 }}}   \\
&=& 2\frac{\pi }
{4}\left( {1 - \left| \delta  \right|} \right)\ln \frac{{1 + \left( {1 - \left| \delta  \right|} \right)^{1/2} }}
{{1 - \left( {1 - \left| \delta  \right|} \right)^{1/2} }} + 2\left[ { - \left( {1 - \left| \delta  \right|} \right)^{1/2}
+ \sqrt {\left| \delta  \right|} \cos ^{ - 1} \sqrt {\left| \delta  \right|} } \right]  \\
  &&\;\;\;\;\;\;\;\;\;\;\;\;\;\;\;\;\;\;\;\;\;\;\;\;\;\;\;\;\;\;\;\;\;\;\;\;\;\;\;\;\;\;\;\;\;\;\;\;\;\;\;\;\; + 2\left( {1 - \left| \delta  \right|} \right)^{1/2} \int_0^{\left( {1
  - \left| \delta  \right|} \right)^{1/2} } {dx\frac{{\sqrt {\left| \delta  \right| - 1} \sinh ^{ - 1} \left( {\frac{x}
{{\sqrt {\left| \delta  \right| - 1} }}} \right)}}
{{ - 1 + x^2 }}}    \label{ironpnictides:d3z21}
\eea
where $\left( \begin{array}{l}
 n \\
 m \\
 \end{array} \right)\equiv \frac{{\Gamma \left( {n + 1} \right)}}
{{\Gamma \left( {m + 1} \right)\Gamma \left( {n - m + 1} \right)}}$ is the binomial coefficient. But
\bea
&&  2\left( {1 - \left| \delta  \right|} \right)^{1/2} \int_0^{\left( {1 - \left| \delta  \right|} \right)^{1/2} } {dx\frac{{\sqrt {\left| \delta  \right| - 1}
 \sinh ^{ - 1} \left( {\frac{x}
{{\sqrt {\left| \delta  \right| - 1} }}} \right)}}
{{ - 1 + x^2 }}}  \\
&=& 2\left( {1 - \left| \delta  \right|} \right)^{1/2} \left\{ {\left. {\sqrt {\left| \delta  \right| - 1} \left( { - \tanh ^{ - 1} x} \right)\sinh ^{ - 1}
 \left( {\frac{x}
{{\sqrt {\left| \delta  \right| - 1} }}} \right)} \right|_{x = 0}^{\left( {1 - \left| \delta  \right|} \right)^{1/2} }
 + \int_0^{\left( {1 - \left| \delta  \right|} \right)^{1/2} } {dx\frac{{\tanh ^{ - 1} x}}
{{\sqrt {1 - x^2 /(1 - \left| \delta  \right|)} }}} } \right\}  \\
&=&  - 2\frac{\pi }
{4}\left( {1 - \left| \delta  \right|} \right)\ln \frac{{1 + \left( {1 - \left| \delta  \right|} \right)^{1/2} }}
{{1 - \left( {1 - \left| \delta  \right|} \right)^{1/2} }} + 2\left( {1 - \left| \delta  \right|} \right)^{1/2} \frac{i}
{4}\left( {1 - \left| \delta  \right|} \right)^{1/2} \left\{ {\pi ^2  - 4\cosh ^{ - 1} \sqrt {\left| \delta  \right|} \ln \left( {
- i\left( {1 - \sqrt {\left| \delta  \right|} } \right)/\sqrt {1 - \left| \delta  \right|} } \right)} \right. \nonumber \\
 && \;\;\;\;\;\;\;\;\;\;\;\;\;\;\;\;\;\;\;\;\;\;\;\;\;\;\;\;\;\;\;\;\;\;\;\;\;\;\;\;\;\;\;\;\;\;\;\;\;\;\;\;\;\;\;\left. { + 4{\text{Li}}_2 \left( {
  - i\left( {1 - \left| \delta  \right|} \right)^{1/2}  - \sqrt {\left| \delta  \right|} } \right) - 4{\text{Li}}_2 \left( {i\left( {1
   - \left| \delta  \right|} \right)^{1/2}  + \sqrt {\left| \delta  \right|} } \right)} \right\}  \label{ironpnictides:d3z2Ipart}
\eea
where ${\text{Li}}_n (z) = \sum\limits_{k = 1}^\infty  {\frac{{z^k }}
{{k^n }}}$ is the polylogarithm function. Substituting Eq.~(\ref{ironpnictides:d3z2Ipart}) back into
Eq.~(\ref{ironpnictides:d3z21}), we have
\bea
  I = 2\left[ { - \left( {1 - \left| \delta  \right|} \right)^{1/2}  + \sqrt {\left| \delta  \right|} \cos ^{ - 1} \sqrt {\left| \delta
   \right|} } \right] + \frac{i}
{{\text{2}}}\left( {1 - \left| \delta  \right|} \right)\left\{ {\pi ^2  - 4\cosh ^{ - 1} \sqrt {\left| \delta  \right|} \ln
 \left( { - i\left( {1 - \sqrt {\left| \delta  \right|} } \right)/\sqrt {1 - \left| \delta  \right|} } \right)} \right. \nonumber \\
  \;\;\;\;\;\;\;\;\;\;\;\;\;\;\;\;\;\;\;\;\;\;\;\;\;\;\;\;\;\;\;\;\;\;\;\;\;\;\;\;\;\;\;\;\;\;\;\;\;\;\left. { + 4{\text{Li}}_2 \left( { - i\left( {1
  - \left| \delta  \right|} \right)^{1/2}  - \sqrt {\left| \delta  \right|} } \right) - 4{\text{Li}}_2 \left( {i\left( {1
  - \left| \delta  \right|} \right)^{1/2}  + \sqrt {\left| \delta  \right|} } \right)} \right\}
\label{ironpnictides:d3z2I}
\eea
Note Eq.~(\ref{ironpnictides:d3z2Ipart}) is an exact result for the integral $I$ in Eq.~(\ref{ironpnictides:d3z2integrals}).
For simplicity here we only consider the analytic limit at $|\delta|=8|\Delta_I|/\gamma^2 \ll 1$.
Within this limit we can get an expansion series of Eq.~(\ref{ironpnictides:d3z2I}) in the order of $|\delta|$,
\be
I = 2\left\{ { - 1 + 2\alpha _0  - 2\alpha _0 \left| \delta  \right| + \frac{\pi }
{3}\left| \delta  \right|^{3/2}  - \frac{1}
{4}\left| \delta  \right|^2  + O\left( {\left| \delta  \right|^{5/2} } \right)} \right\}
 \label{ironpnictides:d3z2Iresult}
\ee
where $\alpha_0 \approx 0.91596$ is the Catalan number.

Now we calculate the integral $II$ in Eq.~(\ref{ironpnictides:d3z2integrals}), which is straightforward in the limit
of $|\delta|=8|\Delta_I|/\gamma^2 \ll 1$.
\be
II = 2i\int_0^{\sqrt {\Lambda _\gamma  } } {dx\sqrt {x^2  + 1 - \left| \delta  \right|} \ln \frac{{1 - ix}}
{{1 + ix}}}  = 4\int_0^{\sqrt {\Lambda _\gamma  } } {dx\sqrt {x^2  + 1} \tan ^{ - 1} x}  - 2\left| \delta
\right|\int_0^{\sqrt {\Lambda _\gamma  } } {dx\frac{{\tan ^{ - 1} x}}
{{\sqrt {1 + x^2 } }}}  + O\left( {\left| \delta  \right|^2 } \right) \label{ironpnictides:d3z2IIintegrals}
\ee
but
\bea
 && 4\int_0^{\sqrt {\Lambda _\gamma  } } {dx\sqrt {x^2  + 1} \tan ^{ - 1} x} \nonumber  \\
   &=& 4\left\{ {\frac{1}
{2}\left. {\left[ {\sqrt {1 + x^2 } \left( {x\tan ^{ - 1} x - 1} \right) + \tan ^{ - 1} x\ln \frac{{1 - ie^{i\tan ^{ - 1} x} }}
{{1 + ie^{i\tan ^{ - 1} x} }} + i{\text{Li}}_2 \left( { - ie^{i\tan ^{ - 1} x} } \right) - i{\text{Li}}_2 \left( {ie^{i\tan ^{ - 1} x} }
\right)} \right]} \right|_{x = 0}^{\sqrt {\Lambda _\gamma  } } } \right\} \nonumber \\
   &=& \pi \Lambda _\gamma   - 4\sqrt {\Lambda _\gamma  }  + \frac{\pi }
{2}\ln \Lambda _\gamma   + \frac{\pi }
{2}\left( {1 + 2\ln 2} \right) + 2 - 4\alpha _0  + O\left( {\frac{1}
{{\sqrt {\Lambda _\gamma  } }}} \right)   \label{ironpnictides:d3z2IIintegral1}
\eea
and
\bea
&&  \int_0^{\sqrt {\Lambda _\gamma  } } {dx\frac{{\tan ^{ - 1} x}}
{{\sqrt {x^2  + 1} }}}  = \left. {\left[ {\tan ^{ - 1} x\ln \frac{{1 - ie^{i\tan ^{ - 1} x} }}
{{1 + ie^{i\tan ^{ - 1} x} }} + i{\text{Li}}_2 \left( { - ie^{i\tan ^{ - 1} x} } \right) - i{\text{Li}}_2 \left( {ie^{i\tan ^{ - 1} x} }
 \right)} \right]} \right|_{x = 0}^{\sqrt {\Lambda _\gamma  } } \nonumber\\
& =& \frac{\pi }
{2}\ln 2 - 2\alpha _0  + \frac{\pi }
{4}\ln \Lambda _\gamma   + O\left( {\frac{1}
{{\sqrt {\Lambda _\gamma  } }}} \right)  \label{ironpnictides:d3z2IIintegral2}
\eea
Substituting the results of Eqs.~(\ref{ironpnictides:d3z2IIintegral1},\ref{ironpnictides:d3z2IIintegral2})
into Eq.~(\ref{ironpnictides:d3z2IIintegrals}), we have
\be
II = \pi \Lambda _\gamma   - 4\sqrt {\Lambda _\gamma  }  + \frac{\pi }
{2}\ln \Lambda _\gamma   + \frac{\pi }
{2}\left( {1 + 2\ln 2} \right) + 2 - 4\alpha _0  - 2\left| \delta  \right|\left( {\frac{\pi }
{2}\ln 2 - 2\alpha _0  + \frac{\pi }
{4}\ln \Lambda _\gamma  } \right) + O\left( {\frac{1}
{{\sqrt {\Lambda _\gamma  } }}} \right) + O(|\delta |^2 )  \label{ironpnictides:d3z2IIresult}
\ee
Combining the results in Eqs.~(\ref{ironpnictides:d3z2Iresult},\ref{ironpnictides:d3z2IIresult}), we finally have
\be
I + II = \pi \Lambda _\gamma   - 4\sqrt {\Lambda _\gamma  }  + \frac{\pi }
{2}\ln \Lambda _\gamma   + \frac{\pi }
{2}\left( {1 + 2\ln 2} \right) - \left( {\pi \ln 2 + \frac{\pi }
{2}\ln \Lambda _\gamma  } \right)\left| \delta  \right| + \frac{\pi }
{3}\left| \delta  \right|^{3/2}  + O\left( {\frac{1}
{{\sqrt {\Lambda _\gamma  } }}} \right) + O(|\delta |^2 ) \label{ironpnictides:d3z2integralresults}
\ee
Substituting Eq.~(\ref{ironpnictides:d3z2integralresults}) into Eq.~(\ref{ironpnictides:v5}), we have
\bea
  \frac{{\alpha_c \Delta _I }}
{{ u_I \gamma^2 }}
   = \frac{{\alpha_c \Delta _I }}
{{(2u_1  + u_2)\gamma^2 }} + \kappa _{\text{0}}  - \kappa _{\text{1}} \left| \delta  \right| + \frac{{\pi }}
{3}|\delta |^{3/2}  + O(\delta ^2 )
\eea
with $\alpha _c  = 64\pi ^3 c^{1/2} \sqrt {c^2  - v^2 }$, and
\be
\kappa _{\text{0}}(w)  = \frac{\alpha_c r(w)}
{{(2u_1  + u_2)\gamma^2 }} +
\pi \Lambda _\gamma   - 4\sqrt {\Lambda _\gamma  }  + \frac{\pi }
{2}\ln \Lambda _\gamma   + \frac{\pi }
{2}\left( {1 + 2\ln 2} \right); \kappa _1  = \pi \ln 2 + \frac{\pi }{2}\ln \Lambda _\gamma =  \pi \ln 2 + \frac{\pi }
{2}\ln\frac{4 }{\Gamma^2} \label{ironpnictides:kappa}
\ee
Then we have
\be
B(|\delta|) \equiv a_{d3z2} \left| \delta  \right| - \frac{{2\pi }}
{3}\left| \delta  \right|^{3/2}  = \kappa _{\text{0}} (w) \label{ironpnictides:d3z2variational}
\ee
with
\be
a_{d3z2}  =  - \frac{{\alpha _c }}
{4}\left( {\frac{1}
{{u_I }} - \frac{1}
{{2u_1  + u_2 }}} \right) + 2\kappa _1  =  - \frac{{16\pi ^3 }}
{{a_c }}\left( {\frac{{c^{3/2} }}
{{u_I }} - \frac{{c^{3/2} }}
{{2u_1  + u_2 }}} \right) + 2\kappa _1 \label{ironpnictides:d3z20}
\ee
From Eq.~(\ref{ironpnictides:d3z2variational}) we can see that the sign of $a_{d3z2}$ will determine the order(s) of the phase transition.
 If $a_{d3z2} < 0$, there is no first order transition since Eq.~(\ref{ironpnictides:d3z2variational}) always has only one solution;
  the Ising order parameter will continuously go to zero as we increase the controlled parameter $w$ in $\kappa_0(w)$.
 Fig.~\ref{fig:sketchprocesscombine}(b) illustrates the process for the phase transitions at $a_{d3z2} < 0$.

If $a_{d3z2} > 0$, a first order transition can happen, since two solutions of Eq.~(\ref{ironpnictides:d3z2variational})
 emerge when $\kappa_0(w) > 0$. From the LHS of Eq.~(\ref{ironpnictides:d3z2variational}),  we can determine that
 $a_{d3z2} > 0$ can happen either at $v \approx c$ (i.e., extreme anisotropy) or at extremely small damping rate.
 In the former case the system is effectively reduced back to the 3D problem, where we roughly recover the 2D results.
For the latter case, it is equivalent to changing the effective dimension $d+z= d+2$ to $d+1$. Therefore in both
of these two extreme situations, the effective dimension of the system becomes 4;
 the Ising coupling ``$-u_I$'' is again marginal, and a first-order
transition is to be expected from RG-based considerations. For the problem we are considering, neither case applies.

For the summation in Eq.~(\ref{ironpnictides:v4}), a similar calculation can be carried out. One can easily find that it is
$\delta$-independent, which is just equal to the $\delta$-independent part of the summation in Eq.~(\ref{ironpnictides:v5}).
Therefore after summing Eq.~(\ref{ironpnictides:v4}) and Eq.~(\ref{ironpnictides:v5}) we will get,
\be
\frac{{2\Delta _I }}
{{u_I }} =  - 2\sigma ^2  + \frac{{\gamma ^2 }}
{{\alpha _c }}\left( { - \kappa _1 \left| \delta  \right| + \frac{\pi }
{3}\left| \delta  \right|^{3/2}  + O\left( {\frac{1}
{{\sqrt {\Lambda _\gamma  } }}} \right) + O\left( {\left| \delta  \right|^2 } \right)} \right)
\ee
i.e.,
\be
\sigma _0  = \frac{\Gamma }
{{8\pi }}\sqrt {\frac{{a_c }}
{\pi }} \sqrt {\frac{1}
{2}\left( {\frac{{16\pi ^3 }}
{{a_c }}\frac{{c^{3/2} }}
{{u_I }} - \kappa _1 } \right)\left| \delta  \right| + \frac{\pi }
{6}\left| \delta  \right|^{3/2} } \label{ironpnictides:d3z2m}
\ee
\begin{figure}[t]
\begin{center}
\includegraphics[width=15.5cm]{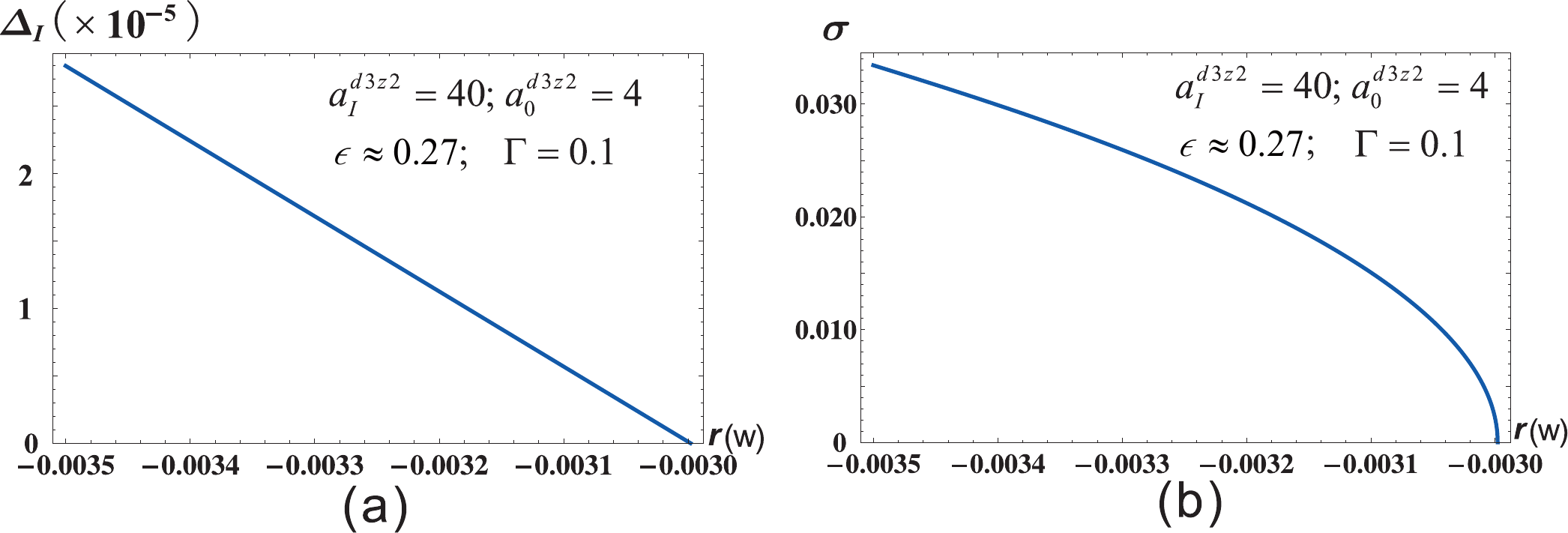}
\end{center}
\caption[phase transition in d3z2 systems when both anisotropy and interaction are weak]{Evolution of (a) the
 Ising order parameter and (b) the magnetic order parameter vs. the control parameter at a moderate strong anisotropy,
 for the 3D case.}
\label{fig:d3z2weak}
\end{figure}
\begin{figure}[t]
\begin{center}
\includegraphics[width=15.5cm]{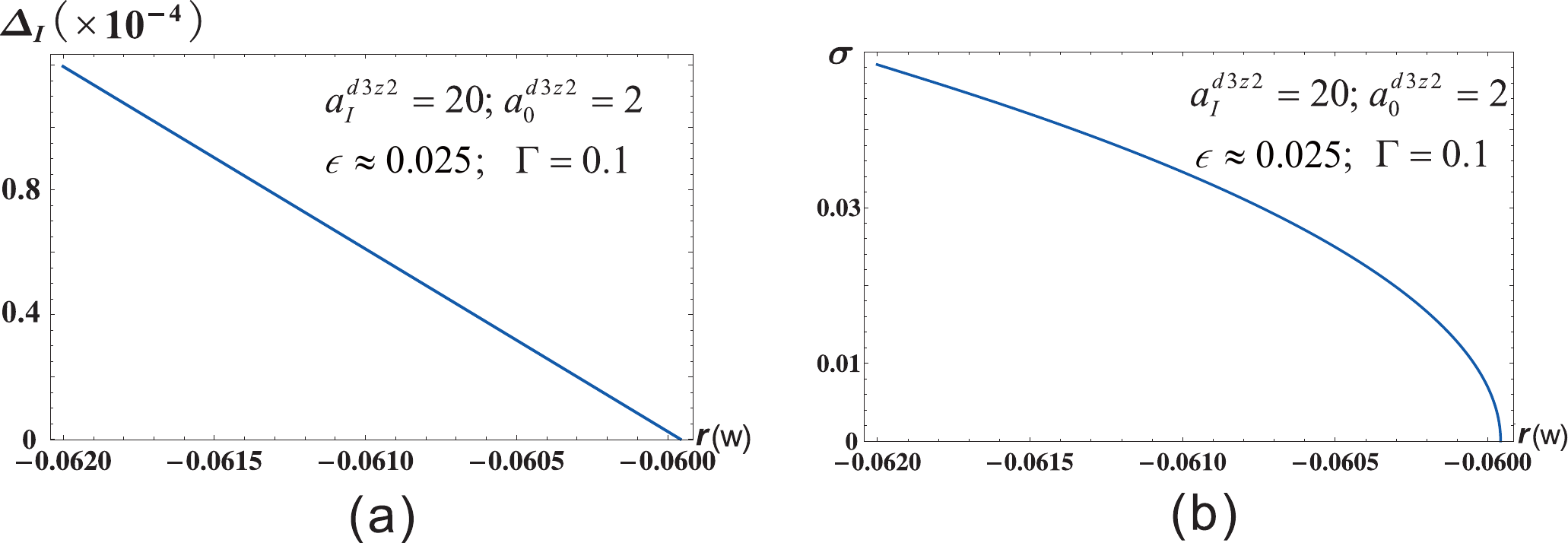}
\end{center}
\caption[phase transition in d3z2 systems when both anisotropy and interaction are strong]{
Evolution of (a) the Ising order parameter and (b) the magnetic order parameter vs. the control parameter
at an extremely strong anisotropy, and with strong interactions, also in the 3D case.}
\label{fig:d3z2strong}
\end{figure}
with dimensionless magnetization $\sigma _0  = c^{1/4} \sigma /\Lambda _c$. From Eq.~(\ref{ironpnictides:d3z2m}) we can see
that when $\delta$ continues to zero, the magnetization will also continuously go to zero, indicating a second-order magnetic
phase transition, and the concurrence of Ising and magnetic phase transitions. From
Eqs.~(\ref{ironpnictides:d3z2variational},\ref{ironpnictides:d3z2m}), we can get Ising order and magnetic order vs.
 the control parameter $r(w)$ in $d=3,\,z=2$ systems
in the limit of $|\delta| = 8|\Delta_I|/\gamma^2 \ll 1$, as shown in Figs.~(\ref{fig:d3z2weak},\ref{fig:d3z2strong}),
where the Ising order $\Delta_I$ and magnetic order $\sigma$ have been respectively re-scaled into dimensionless quantities
via $\Delta_I \to \Delta_I/(c \Lambda_c^2)$ and $\sigma \to c^{1/4}\sigma/\Lambda_c = \sigma_0$ (for convenience we also
introduce a group of dimensionless parameters
$a_I^{d3z2} = c^{3/2}/u_I, a_0^{d3z2} =  c^{3/2}/(2u_1 + u_2), a_c = c/\sqrt{c^2 - v^2}, \Gamma = \gamma/(c^{1/2} \Lambda_c)$).
At moderate strong anisotropy $\epsilon \approx 0.27$ (Fig.~\ref{fig:d3z2weak}), it shows continuous quantum
phase transitions and concurrence of the Ising and magnetic orders when increasing $w$. As in the 2D case we also
study the effect of strong anisotropy at $\epsilon \approx 0.025$ (Fig.~\ref{fig:d3z2strong}),
where the continuous phase transitions persist, and the two transitions are concurrent. This is consistent with the
RG considerations: given that the effective dimensionality in this case is $d+z=5$, the quartic coupling $-u_I$ becomes
irrelevant w.r.t. to the underlying O($3$) transition and will therefore not destabilize the continuous nature of the transition.

\end{document}